 \documentclass[10pt,aps,prfluids,print,superscriptaddress,longbibliography,notitlepage]{revtex4-1}

\usepackage{graphicx}
\usepackage{amsmath,amssymb}
\graphicspath{{figures/}}
\usepackage[export]{adjustbox}
\usepackage{amsmath}
\usepackage{mathtools}
\usepackage{bm}
\usepackage{multirow}
\usepackage{bigints}
\usepackage{xcolor}
\usepackage{tabularx}
\usepackage{enumitem}
\usepackage[%
  colorlinks=true,
  urlcolor=blue,
  linkcolor=blue,
  citecolor=blue
]{hyperref}
\usepackage{mathrsfs}
\usepackage{dblfloatfix}
\usepackage{float}

\def\XXint#1#2#3{{\setbox0=\hbox{$#1{#2#3}{\int}$}
     \vcenter{\hbox{$#2#3$}}\kern-.5\wd0}}

\newcommand{\beq}{\begin{equation}}
\newcommand{\eeq}{\end{equation}}

\definecolor{forestgreen}{RGB}{74,103,65}

\begin{document}

\title{{\textcolor{black}{Stability of
co-annular active and passive confined fluids}}}
\author{Tanumoy Dhar}
\affiliation{Department of Mechanical and Aerospace Engineering, University of California San Diego, \\ 9500 Gilman Drive, La Jolla, CA 92093, USA}
\author{Michael J. Shelley}
\affiliation{Center for Computational Biology, Flatiron Institute, 162 Fifth Avenue, New York, NY 10010, USA}
\affiliation{Courant Institute of Mathematical Sciences, New York University, \\ 251 Mercer Street, New York, NY 10012, USA}
\author{David Saintillan}
\email{dstn@ucsd.edu}
\affiliation{Department of Mechanical and Aerospace Engineering, University of California 
San Diego, \\ 9500 Gilman Drive, La Jolla, CA 92093, USA}
\affiliation{Center for Computational Biology, Flatiron Institute, 162 Fifth Avenue, New York, NY 10010, USA}

\begin{abstract}
The translation and shape deformations of a passive viscous Newtonian droplet immersed in an active nematic liquid crystal under circular confinement are analyzed using a linear stability analysis. We focus on the case of a sharply aligned active nematic in the limit of strong elastic relaxation in two dimensions.\ Using an active liquid crystal model, we employ the Lorentz reciprocal theorem for Stokes flow to study the growth of interfacial perturbations as a result of both active and elastic stresses.\ Instabilities are uncovered in both extensile and contractile systems, for which growth rates are calculated and presented in terms of the dimensionless ratios of active, elastic, and capillary stresses, as well as the viscosity ratio between the two fluids. We also extend our theory to analyze the inverse scenario, namely, the stability of an active nematic droplet surrounded by a passive viscous layer.\ Our results highlight the subtle interplay of capillary, active, elastic, and viscous stresses in governing droplet stability.\ The instabilities uncovered here may be relevant to a plethora of biological active systems, from the dynamics of passive droplets in bacterial suspensions to the organization of subcellular compartments inside the cell and cell nucleus. 
\end{abstract}

\maketitle


\section{Introduction}

The interaction of active materials with passive phases is relevant to many biological systems spanning a range of length scales. Examples on large scales include: front propagation in spreading bacterial swarms and suspensions, where complex interfacial instabilities have been observed \cite{patteson2018propagation,martinez2022morphological}; cell migration during the formation of cancer metastasis \cite{nurnberg2011nucleating,hoshino2013signaling}; or the migration of confluent epithelial cell layers during wound healing \cite{martin2004parallels,armengol2023epithelia}. On cellular length scales, the organization of various subcellular compartments and organelles hinges on their interaction with the active cytoskeleton and cytoplasmic flows \cite{chu2017origin,mogilner2018intracellular}. On yet smaller scales, the placement and structural organization of subnuclear bodies and heterochromatin compartments inside the cell nucleus have been hypothesized to be affected by the dynamics of transcriptionally active euchromatin \cite{caragine2018surface,mahajan2022euchromatin,rautu2015hetero}. Interfaces between active and passive phases have also been of interest in synthetic and reconstituted systems, e.g., in active nematic fluids composed of microtubules and molecular motors \cite{sanchez2012spontaneous,shelley2016dynamics}, where activity has been shown to destabilize interfaces and drive shape fluctuations and active waves \cite{adkins2022dynamics}. 

In all of these systems, the dynamics and emerging morphologies of active/passive interfaces stem from the interplay of active, elastic, viscous, and \textcolor{black}{capillary stresses}. A perturbation to the interfacial shape affects the orientation of the active matter constituents, thus giving rise to both active and elastic stresses. These stresses in turn result in tractions on the interface as well as flows in the material, whose net effect can be either stabilizing or destabilizing. A few theoretical and computational models have been proposed in recent years to explain these effects. The dynamics of active droplets, either in free space or near boundaries, has been studied numerically and analytically, where active stresses and the flows they induce can lead to spontaneous propulsion, fingering instabilities, division or spreading \cite{tjhung2012spontaneous,joanny2012drop,giomi2014spontaneous,whitfield2014active,khoromskaia2015motility,young2021many,alert2022fingering}. Active films have also been considered, where interfacial instabilities can either be triggered or stabilized by activity depending on the sign of active stresses \cite{voituriez2006generic,sankararaman2009instabilities,blow2017motility,alonso2019interfacial}. Other studies have analyzed the motion of passive objects, either rigid or deformable, suspended in active suspensions, where active stresses can give rise to spontaneous transport \cite{freund2023object,freund2024free}, rotation \cite{furthauer2012taylor}, and deformations \cite{chandler2024active}. 

In the present work, we analyze a canonical system consisting of a passive viscous Newtonian droplet immersed in an active nematic liquid crystal under circular confinement. The opposite scenario of an active nematic droplet surrounded by a passive viscous layer is also addressed. We consider the sharply aligned limit for the nematic, and use strong anchoring boundary conditions at the interface and domain boundaries, a regime where the configuration of the nematic field is entirely governed by the system's geometry. Using a linear stability analysis along with the Lorentz reciprocal theorem \cite{masoud2019reciprocal}, we study the stability of the drop under both translation and deformation, and calculate growth rates in terms of the governing parameters. We uncover instabilities in both extensile and contractile systems, which may be relevant to a wide range of biological active systems, from the dynamics of passive droplets in bacterial suspensions to the organization of subcellular compartments inside the cell and cell nucleus. 

The paper is organized as follows. We formulate the problem in Sec.~\ref{section2}, where we present the governing equations, boundary conditions, and non-dimensionalization.\ Section~\ref{section3} presents details of the linear stability analysis of the axisymmetric base state and a discussion of the dispersion relation and growth rates.\ The inverted problem, i.e., the stability of an active nematic drop within a passive layer and its associated dispersion relation, is discussed in Sec.~\ref{section4}. We conclude in Sec.~\ref{section5}.

\textcolor{black}{\section{System formulation}\label{section2}}

\begin{figure}[t]
\includegraphics[width=0.9\textwidth]{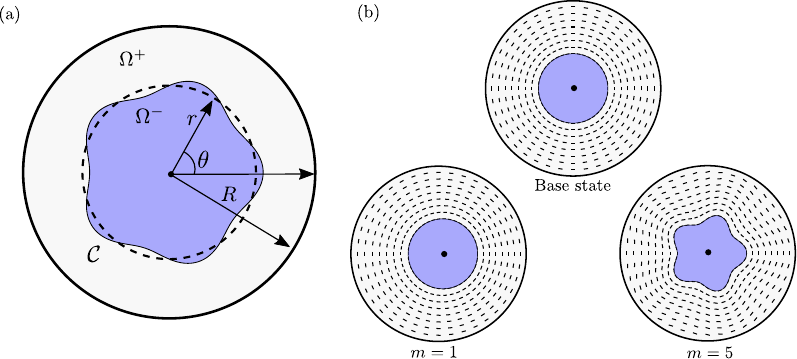}
\caption{Problem definition: a viscous drop is placed in an apolar active nematic suspension under circular confinement, with strong tangential anchoring at both the drop surface and outer boundary (a) (see more details on the symbols in the main text). Figure (b) shows the configuration of the nematic director (black segments) in the undisturbed base state and under perturbations with modes $m=1$ and $m=5$.}
\label{schematic}
\end{figure}

We analyze the stability of a viscous Newtonian droplet surrounded by a sharply aligned active nematic fluid in circular confinement in two dimensions (Fig.~\ref{schematic}).\ The system is enclosed by a circular boundary of radius $R$ centered at the origin $O$. In polar coordinates $(r,\theta)$ \textcolor{black}{with unit basis vectors $\smash{(\hat{\mathbf{r}},\hat{\boldsymbol{\theta}})}$}, the shape of the interface between the drop and nematic layer is described by the Monge representation $r=\eta R+\zeta(\theta,t)$, where $\eta\in(0,1)$ is the dimensionless radius ratio, and $\zeta$ captures deviations from the reference base state where the drop is circular and centered at the origin. We denote by $\Omega^-$ the domain occupied by the passive drop, by $\Omega^+$ the domain occupied by the nematic, and by $\mathcal{C}$ the curve describing the shape of the interface between them.\\

\subsection{Nematodynamics}

Orientational order in the active nematic layer surrounding the droplet is described by a nematic director field $\mathbf{q}(\mathbf{x},t)$ (with $|\mathbf{q}|=1$ in the sharply aligned limit), which satisfies \textcolor{black}{the Ericksen--Leslie model} \cite{ericksen1961conservation,ericksen1962hydrostatic,leslie1968some,LLKP1986,degennes1995physics,S2004,gao2017analytical} 
\begin{align}\label{eq:nematodynamics1}
\frac{\partial \mathbf{q}}{\partial t} +\mathbf{u}\cdot \mathbf{\nabla q} = (\mathbf{I}-\mathbf{qq})\cdot( \varpi \mathbf{e} - \mathbf{w})\cdot\mathbf{q} + \frac{1}{\Gamma}(\mathbf{I}-\mathbf{qq})\cdot \mathbf{h},
\end{align}
where $\mathbf{u}$ is the fluid velocity in the nematic. The rate-of-strain tensor $\mathbf{e}=(\nabla\mathbf{u}+\nabla\mathbf{u}^\mathrm{T})/2$ and rate-of-rotation tensor $\mathbf{w}=(\nabla\mathbf{u}-\nabla\mathbf{u}^\mathrm{T})/2$ depend on the velocity gradient defined as $(\nabla\mathbf{u})_{ij}=\partial_i u_j$, and $\varpi$ denotes the flow alignment parameter. The last term in Eq.~(\ref{eq:nematodynamics1}) captures nematic elasticity, and involves the rotational viscosity $\Gamma$ and molecular field $\mathbf{h}$, which derives from a free energy density.\
Under the commonly used one-constant approximation \cite{degennes1995physics}, we express the free energy density $\mathcal{F}_d$ as
\begin{align}
\mathcal{F}_d=\frac{K}{2}\| \nabla \mathbf{q}\|^2,
\end{align}
from which the molecular field is obtained as
\begin{align}
\mathbf{h}=-\frac{\delta \mathcal{F}_d}{\delta \mathbf{q}}=K\nabla^2\mathbf{q}.
\end{align}
Nematic elasticity thus enters Eq.~(\ref{eq:nematodynamics1}) as a diffusive term, with the ratio $K/\Gamma$ playing the role of a diffusivity. The nematic is assumed to satisfy strong parallel anchoring conditions at both the outer domain boundary and droplet interface. In two dimensions, the director field is equivalently described by a single angle $\beta(\mathbf{x},t)$ such that $\mathbf{q}=[\cos\beta,\sin\beta]$; we make further use of this representation below. 

\subsection{Mass and momentum balance}

Conservation of mass and momentum in the active nematic read
\begin{align}
\nabla\cdot\mathbf{u}=0, \qquad \nabla \cdot (\boldsymbol{\sigma}^h+\boldsymbol{\sigma}^a+\boldsymbol{\sigma}^e)=\boldsymbol{0},   \label{eq:stokes1}
\end{align}
where $\boldsymbol{\sigma}^h$, $\boldsymbol{\sigma}^a$ and $\boldsymbol{\sigma}^e$ are the hydrodynamic, active, and elastic (or Ericksen--Leslie) stress tensors, respectively:
\begin{align}
\boldsymbol{\sigma}^h &= -p\,\mathbf{I}+2\mu\,\mathbf{e},  \\  
\boldsymbol{\sigma}^a & = -\alpha \mathbf{qq},  \label{eq:sigmaa} \\
\begin{split}
    \textcolor{black}{\boldsymbol{\sigma}^e} & \,\,\textcolor{black}{= \big(\tfrac{1}{2}K \|\nabla \mathbf{q}\|^2-\varpi \Gamma \mathbf{pp}:\mathbf{e} \big)\mathbf{I}-K \nabla\mathbf{q}\cdot\nabla\mathbf{q}^\mathrm{T} +\varpi^2\Gamma (\mathbf{qq}\,:\,\mathbf{e})\mathbf{qq}} \\
    &\,\,\,\,\,\,\,\,\textcolor{black}{- \tfrac{1}{2}K\left[(\varpi+1)\mathbf{q}(\nabla^2\mathbf{q})+ (\varpi-1)(\nabla^2\mathbf{q})\mathbf{q} + 2\varpi \| \nabla \mathbf{q}\|^2 \mathbf{qq}\right].}
    \label{eq:sigmae}
    \end{split}
\end{align}
Here, $p$ is the fluid pressure, and $\mu$ is the shear viscosity of the fluid. The active stress magnitude is $\alpha=n\varsigma_a$, where $n$ is the number density of active dipoles and $\varsigma_a$ is the dipole strength or stresslet; $\alpha>0$ for an extensile nematic, and $\alpha<0$ in the contractile case \cite{saintillan2013active,saintillan2018rheology}. \textcolor{black}{We note that several expressions for $\boldsymbol{\sigma}^e$ with varying levels of complexity exist in the literature \cite{LLKP1986,degennes1995physics,S2004}. The version provided in Eq.~(\ref{eq:sigmae}) assumes an isotropic viscous response; see Cates \& Tjhung \cite{CT2018} for a derivation. In Eq.~(\ref{eq:stokes1}), we adopt the convention $(\nabla\cdot \boldsymbol{\sigma})_{i}=\partial_k \sigma_{ki}$ for the tensorial divergence; accordingly, the traction on a surface with outward unit normal $\mathbf{n}$ is defined as $\mathbf{t}=\mathbf{n}\cdot\boldsymbol{\sigma}$. }  

In the passive viscous drop, the only stress is hydrodynamic, and the governing equations are
\begin{align}
\nabla\cdot\overline{\mathbf{u}}=0, \qquad \nabla \cdot \overline{\boldsymbol{\sigma}}^h=\boldsymbol{0},  \label{eq:stokes2}
\end{align}
where $\overline{\boldsymbol{\sigma}}^h = -\overline{p}\,\mathbf{I}+2\overline{\mu}\,\overline{\mathbf{e}}$ and $\overline{\mathbf{e}}=\tfrac{1}{2}(\nabla \overline{\mathbf{u}}+ \nabla\overline{\mathbf{u}}^{\mathrm{T}})$. Here and in the following, variables associated with flow inside the drop are decorated with an overbar. 

\subsection{Boundary conditions}

At the outer boundary $r=R$, the no-slip condition $\mathbf{u}=\mathbf{0}$ applies. There, the nematic director is tangent to the boundary: $\mathbf{q}\cdot\hat{\mathbf{r}}=0$.

\textcolor{black}{To parametrize the shape of the drop, we introduce the function $F(r,\theta,t)= r-\eta R -\zeta(\theta,t)$ so that the interface $\mathcal{C}$ between the two fluids is defined as $F=0$. Consequently, the outward unit normal to the interface is}
\begin{align}
\mathbf{n}=\frac{\nabla F}{|\nabla F|}.
\end{align}
The kinematic boundary condition describing the evolution of the interface is
\begin{align}
\frac{\partial F}{\partial t}+\mathbf{u}\cdot \nabla F=0,
\end{align}
i.e.,
\begin{align}
\frac{\partial \zeta}{\partial t}=u_r-\frac{u_\theta}{r}\frac{\partial \zeta}{\partial \theta}. \label{eq:kinematicBC}
\end{align}
Finally, the fluid velocity is continuous at the interface: $\mathbf{u}=\overline{\mathbf{u}}$ at $F=0$. 

\textcolor{black}{Neglecting Marangoni effects, the dynamic boundary condition is given by the stress balance
\begin{equation}
\mathbf{n}\cdot(\boldsymbol{\sigma}^h+\boldsymbol{\sigma}^a+\boldsymbol{\sigma}^e-\overline{\boldsymbol{\sigma}}^h)+\mathbf{t}^{e}_{s}=\gamma \kappa \mathbf{n},  \label{eq:dynamicBC}
\end{equation}
where $\gamma$ is the uniform surface tension, and $\kappa = \nabla\cdot \mathbf{n}$ is the local curvature. The parallel anchoring condition for the nematic at the drop surface reads $\mathbf{q}\cdot \mathbf{n}=0$ at $F=0$, and results in an additional elastic surface traction $\mathbf{t}^{e}_{s}$ in Eq.~(\ref{eq:dynamicBC}) expressed as \cite{CS2024}:  \vspace{-0.2cm}
\begin{equation}
    \mathbf{t}^{e}_{s} = K\frac{\partial }{\partial s}\bigg(\frac{\partial \beta}{\partial n}\mathbf{n}\bigg).  \label{eq:elastictraction}
\end{equation}Here, $\partial/\partial n$ and $\partial/\partial s$ denote the normal and arc-length derivatives along the interface, respectively, with the normal pointing into the nematic and the direction of arclength determined by a ninety-degree counterclockwise rotation of the normal.}

\textcolor{black}{The choice of strong anchoring boundary conditions at the domain boundaries is convenient mathematically as it simplifies the solution of the boundary problem for $\beta$. The choice of parallel versus normal anchoring, however, is the most physically relevant as many of the systems that motivate this work (such as microtubule suspensions) involve elongated molecules that spontaneous align with boundaries as a result of steric interactions. Our analysis could be extended to allow for weak anchoring, with the introduction of an additional physical constant capturing the strength of anchoring \cite{CS2024}. }

\subsection{Non-dimensionalization}

We scale the equations using length scale $R$, time scale $\mu/|\alpha|$, and velocity scale $|\alpha| R/\mu$. The nematodynamic equation (\ref{eq:nematodynamics1}) becomes
\begin{align}\label{eq:nematodynamics2}
\frac{\partial \mathbf{q}}{\partial t} +\mathbf{u}\cdot \mathbf{\nabla q} = (\mathbf{I}-\mathbf{qq})\cdot( \varpi \mathbf{e} - \mathbf{w})\cdot\mathbf{q} + \frac{1}{\mathrm{Pe}}(\mathbf{I}-\mathbf{qq})\cdot \nabla^2 \mathbf{q},
\end{align}
where the active P\'eclet number $\mathrm{Pe}=\Gamma |\alpha|R^2/K\mu$ compares the effects of the fluid flow to nematic relaxation on the dynamics of $\mathbf{q}$. 

In the rest of the paper, we will assume $\mathrm{Pe}\to 0$ (strong nematic relaxation), in which case the equation for $\mathbf{q}$ simplifies to
\begin{align}
(\mathbf{I}-\mathbf{q}\mathbf{q})\cdot\nabla^2\mathbf{q}=\mathbf{0} \qquad \implies \qquad \nabla^2 \beta =0.   \label{eq:laplacebeta}
\end{align}
\textcolor{black}{In this limit, the configuration of the nematic is thus purely a boundary value problem governed by the geometry of the nematic layer through the anchoring boundary conditions. The only effect of the flow on $\mathbf{q}$ is through the kinematic boundary condition (\ref{eq:kinematicBC}), which governs the shape of the interface.}


In the nematic, we use $|\alpha|$ to scale the pressure and stress. The stress tensors inside the active nematic then become
\begin{equation}
\boldsymbol{\sigma}^h = -p\,\mathbf{I}+2\,\mathbf{e},  \qquad 
\boldsymbol{\sigma}^a  = -S \mathbf{qq}, \qquad
\textcolor{black}{\boldsymbol{\sigma}^e  =- \xi^2 \nabla\mathbf{q}\cdot \nabla \mathbf{q}^\mathrm{T},} 
  \label{eq:dimlessstress}
\end{equation}\textcolor{black}{Note that the elastic stress was simplified under that assumption of $\mathrm{Pe}\rightarrow 0$ and by making use of Eq.~(\ref{eq:laplacebeta}), and that its isotropic part is henceforth omitted as it can absorbed in the fluid pressure.} 
Inside the passive drop, the hydrodynamic stress reads
\begin{equation}
\overline{\boldsymbol{\sigma}}^h = -\overline{p}\,\mathbf{I}+2\lambda \,\overline{\mathbf{e}}.
\end{equation}
where $\lambda=\overline{\mu}/\mu$ is the viscosity ratio. \textcolor{black}{The active stress coefficient in Eq.~(\ref{eq:dimlessstress}) is $S = \alpha/|\alpha|$, with $S=+1$ for an extensile nematic and $S=-1$ for a contractile nematic.} The relative magnitude of elastic and active stresses is captured by the dimensionless parameter $\xi=\sqrt{K/|\alpha|}/R$, or ratio of the active length scale $\sqrt{K/|\alpha|}$ to the system size. \textcolor{black}{The parameter $\xi$ is also related to the Ericksen number used in the literature on liquid crystals as $\mathrm{Er}=1/\xi^2$ \cite{CL2000,chandler2024active}.} Active stresses are negligible when $\xi\gg 1$, whereas they dominate when \textcolor{black}{$\xi\to 0$}, a regime where active turbulence is typically observed. In the present work, we assume \textcolor{black}{that $\xi$ is of order unity}.

The kinematic boundary condition (\ref{eq:kinematicBC}) remains unchanged under non-dimensionalization, and the stress balance (\ref{eq:dynamicBC}) becomes
\begin{equation}
\textcolor{black}{\mathbf{n}\cdot(\boldsymbol{\sigma}^h+\boldsymbol{\sigma}^a+\boldsymbol{\sigma}^e-\overline{\boldsymbol{\sigma}}^h)+\mathbf{t}^e_s=\frac{1}{\mathrm{Ca}} \kappa  \mathbf{n}, }  \label{eq:dynBC}
\end{equation}\textcolor{black}{where $\mathbf{t}^e_s$ is given by Eq.~(\ref{eq:elastictraction}) with $K$ replaced with $\xi^2$, and} where we have introduced the active capillary number $\mathrm{Ca}=|\alpha| R/\gamma$ comparing the magnitudes of active \textcolor{black}{versus} capillary stresses. Under the assumption of vanishing P\'eclet number, the system is entirely governed by five parameters: the radius ratio $\eta$, \textcolor{black}{the sign coefficient $S=\pm 1$ of the active stress}, the capillary number $\mathrm{Ca}$, the dimensionless elastic constant $\xi$, and the viscosity ratio~$\lambda$.

\section{Linear stability analysis}\label{section3}

We analyze the stability of small perturbations from the stable case of a circular drop centered at the origin. In the following, subscripts 0 and 1 are used to denote the base state and linear perturbations, respectively. 

\subsection{Circular base state}

In the unperturbed state where the drop is circular and centered at the origin ($\zeta_0=0$), the nematic field lines are concentric circles with
\textcolor{black}{\begin{equation}
\beta_0=\theta+\frac{\pi}{2}, \label{eq:beta0}
\end{equation}}which satisfies Eq.~(\ref{eq:laplacebeta}) along with the anchoring boundary conditions at $r=\eta$ and $1$.\ The corresponding nematic director is $\smash{\mathbf{q}_0=\hat{\boldsymbol{\theta}}}$. Both fluids are still ($\mathbf{u}_0=\overline{\mathbf{u}}_0=\mathbf{0}$). The base-state active and elastic stress tensors in the outer \textcolor{black}{nematic} layer are given by
\begin{equation}
\boldsymbol{\sigma}^a_0=-S\mathbf{q}_0 \mathbf{q}_0=-S \hat{\boldsymbol{\theta}}\hat{\boldsymbol{\theta}}, \qquad  \textcolor{black}{\boldsymbol{\sigma}^e_0=-\xi^2 \nabla\mathbf{q}_0\cdot \nabla \mathbf{q}_0^\mathrm{T} = -\frac{\xi^2}{r^2}\hat{\boldsymbol{\theta}}\hat{\boldsymbol{\theta}}.   }\label{eq:basestress}
\end{equation}They both induce a pressure gradient in the \textcolor{black}{nematic} fluid:
\begin{equation}
\nabla p_0 = \nabla\cdot (\boldsymbol{\sigma}^a_0+\boldsymbol{\sigma}^e_0)=
\left(\frac{S}{r}+\frac{\xi^2}{r^3}\right)\hat{\mathbf{r}}=  \nabla \left(S\ln r - \frac{\xi^{2}}{2r^{2}}\right),
\end{equation}
from which 
\begin{equation}
p_0(r) = S\ln r - \frac{\xi^{2}}{2r^{2}}.   \label{eq:basepressurenematic}
\end{equation}

The pressure inside the passive drop is uniform and is dictated by the stress balance (\ref{eq:dynBC}):
\textcolor{black}{\begin{equation}
\overline{p}_0 = S\ln \eta - \frac{\xi^{2}}{2\eta^{2}} + \frac{1}{\eta \mathrm{Ca}}.   \label{eq:basepressure}
\end{equation}}

\subsection{Domain perturbation and linearized nematic field}

\textcolor{black}{Leveraging periodicity in the $\theta$ direction}, we perturb the shape of the interface \textcolor{black}{with an azimuthal Fourier} mode 
\begin{equation}
\zeta(\theta,t) = \epsilon \zeta_1(\theta,t)= \epsilon \widetilde{\zeta}_1(t) \cos m\theta,  \label{eq:shapedef}
\end{equation}
where $|\epsilon|\ll 1$ and $m\in \mathbb{N}$. We assume $m$ to be given, and our goal is to analyze the growth of the perturbation amplitude $\widetilde{\zeta}_1(t)$. At linear order, the perturbed outward normal and curvature along the interface are, respectively,
\begin{align}
\mathbf{n} & = \mathbf{n}_0+\epsilon \mathbf{n}_1 +O(\epsilon^2) = \hat{\mathbf{r}} + \epsilon \widetilde{\zeta}_1 \frac{m}{\eta} \sin m\theta \,\hat{\boldsymbol{\theta}} + O(\epsilon^2), \\
\kappa & = \kappa_0 + \epsilon \kappa_1 +O(\epsilon^2)=\frac{1}{\eta}+\epsilon \widetilde{\zeta}_1 \frac{m^2-1}{\eta^2}\cos m \theta +O(\epsilon^2).
\end{align}

The instantaneous shape of the interface fully governs that of the nematic field lines. The director angle is expanded as $\beta(r,\theta) = \beta_0 + \epsilon \beta_1+O(\epsilon^2)$. The boundary conditions on the perturbation angle are $
\beta_1=0$ at $r=1$, and 
\begin{equation}
\beta_1 = \widetilde{\zeta}_1 \frac{m}{\eta}\sin m \theta \quad \mathrm{at}\quad r = \eta.   \label{eq:beta1bc}
\end{equation} 
A solution of Laplace's equation $\nabla^2\beta_1=0$ that satisfies both of these conditions is readily obtained by separation of variables,
\begin{equation}
\beta_1(r,\theta) = \widetilde{\zeta}_1 \frac{m \eta^{m-1}}{\eta^{2m}-1}(r^m-r^{-m})\sin m \theta, 
\end{equation}
and the corresponding expansion for the nematic director follows as
\begin{equation}
\mathbf{q}=\hat{\boldsymbol{\theta}}-\epsilon \beta_1 \hat{\mathbf{r}}+O(\epsilon^2).   \label{eq:linq}
\end{equation}
Typical director fields corresponding to $m=1$ and $5$ are illustrated in Fig.~\ref{schematic}(b,c).

\subsection{Linearized active and elastic stresses}

The linearization of active and elastic stresses is obtained by inserting Eq.~(\ref{eq:linq}) into Eq.~(\ref{eq:dimlessstress}). Starting with the active stress,
\begin{equation}
\boldsymbol{\sigma}^a =\boldsymbol{\sigma}^a_0+ \epsilon \boldsymbol{\sigma}^a_1+O(\epsilon^2)=  -S \hat{\boldsymbol{\theta}}\hat{\boldsymbol{\theta}}+\epsilon  S \beta_1  \big(\hat{\mathbf{r}}\hat{\boldsymbol{\theta}}+\hat{\boldsymbol{\theta}}\hat{\mathbf{r}}\big)+O(\epsilon^2).   \label{eq:sigmaalin}
\end{equation}
To obtain the elastic stress, we first calculate the gradient of $\mathbf{q}$,
\begin{equation}
\nabla \mathbf{q} = \nabla \hat{\boldsymbol{\theta}}-\epsilon \left[ (\nabla \beta_1) \hat{\mathbf{r}}+ \beta_1 \nabla \hat{\mathbf{r}}\right]+O(\epsilon^2)= -\frac{1}{r}\hat{\boldsymbol{\theta}}\hat{\mathbf{r}}-\epsilon \left[\partial_r \beta_1 \hat{\mathbf{r}}\hat{\mathbf{r}} +\frac{\partial_\theta \beta_1}{r}\hat{\boldsymbol{\theta}}\hat{\mathbf{r}}+\frac{\beta_1}{r} \hat{\boldsymbol{\theta}}\hat{\boldsymbol{\theta}}\right]+O(\epsilon^2).
\end{equation}
The linearized elastic stress then reads
\textcolor{black}{\begin{equation}
\boldsymbol{\sigma}^e =\boldsymbol{\sigma}^e_0+ \epsilon \boldsymbol{\sigma}^e_1+O(\epsilon^2) = -\frac{\xi^2}{r^2}\hat{\boldsymbol{\theta}}\hat{\boldsymbol{\theta}} -\epsilon \frac{\xi^2}{r^2} \left[2\partial_\theta \beta_1\hat{\boldsymbol{\theta}}\hat{\boldsymbol{\theta}} + r\partial_{r}\beta_1 \big(\hat{\mathbf{r}}\hat{\boldsymbol{\theta}}+\hat{\boldsymbol{\theta}}\hat{\mathbf{r}}\big) \right]+O(\epsilon^2).  \label{eq:sigmaelin}
\end{equation}}

\subsection{Linearized flow problem and boundary conditions}

The fluid velocity and pressure perturbations satisfy the Stokes equations (\ref{eq:stokes1}) and (\ref{eq:stokes2}) at linear order in $\epsilon$:
\begin{align}
&\nabla \cdot {\mathbf{u}}_1=0, \qquad \nabla \cdot ( \boldsymbol{\sigma}^h_1 +\boldsymbol{\sigma}^a_1+\boldsymbol{\sigma}^e_1)=\mathbf{0}, &  \mathrm{for} \,\,\, r> \eta+\zeta, \label{eq:order1flow} \\
&\nabla \cdot \overline{\mathbf{u}}_1=0, \qquad \nabla \cdot \overline{\boldsymbol{\sigma}}^h_1 = \mathbf{0}, & \mathrm{for} \,\,\, r< \eta+\zeta, \label{eq:order1flow2}
\end{align}
where the active and elastic stresses in Eq.~(\ref{eq:order1flow}) are known from Eqs.~(\ref{eq:sigmaalin}) and (\ref{eq:sigmaelin}). The flows in the two regions are coupled via the dynamic boundary condition (\ref{eq:dynBC}), which applies at $r=\eta+\zeta$. 

Our method of solution, described in Sec.~\ref{sec:Lorentz}, will rely on the Lorentz reciprocal theorem for Stokes flow \cite{masoud2019reciprocal}. To this end, it is convenient to first recast the flow problem around a circular boundary ($r=\eta$). \textcolor{black}{Specifically, we need to apply the dynamic boundary condition (\ref{eq:dynBC}) on the circle, which is achieved by expanding the relevant fields as Taylor series in the neighborhood of $\zeta=0$.} For example, the stress tensors can be expanded as:
\begin{equation}
\boldsymbol{\sigma}(r=\eta+\epsilon \zeta_1) = \boldsymbol{\sigma}(r=\eta) + \epsilon \zeta_1 \partial_r \boldsymbol{\sigma}(r=\eta) + O(\epsilon^2).   \label{eq:Taylor}
\end{equation}
\textcolor{black}{The elastic surface traction, which arises at order $\epsilon$, is expanded separately as
\begin{equation}
\begin{split}
\mathbf{t}^{e}_{s,1} =   \epsilon\frac{\xi^2}{\eta}\bigg[\left(\partial_{\theta}\partial_{r}\beta_{1}|_{r=\eta}+\widetilde{\zeta_{1}}\frac{m^{2}}{\eta^{2}}\cos m\theta\right) \hat{\mathbf{r}} + \left(\partial_{r}\beta_{1}|_{r=\eta} + \widetilde{\zeta_{1}}\frac{m}{\eta^{2}}\sin m\theta\right)\hat{\boldsymbol{\theta}}  \bigg]+O(\epsilon^2).  \label{eq:lintse}
\end{split}
\end{equation}}Inserting Eqs.~(\ref{eq:Taylor}) and (\ref{eq:lintse}) into Eq.~(\ref{eq:dynBC}), linearizing, and simplifying using Eqs.~(\ref{eq:basestress})--(\ref{eq:basepressure}), we obtain the following dynamic boundary condition at $O(\epsilon)$,
\textcolor{black}{\begin{equation}
\begin{split}
\hat{\mathbf{r}}\cdot \big(\boldsymbol{\sigma}^h_1+\boldsymbol{\sigma}^a_1+\boldsymbol{\sigma}^e_1- \overline{\boldsymbol{\sigma}}^h_1 \big) & =\frac{\kappa_1 \hat{\mathbf{r}}}{\mathrm{Ca}}+\frac{\widetilde{\zeta_1}}{\eta}S\big(\cos m\theta \,\hat{\mathbf{r}}+m\sin m\theta\,\hat{\boldsymbol{\theta}} \big) \\ &- \widetilde{\zeta_1 }\frac{\xi^{2}}{\eta^{3}}\bigg[\bigg(\frac{m^{3}(\eta^{2m}+1)}{\eta^{2m}-1} + m^{2}-1\bigg)\cos m\theta \,\hat{\mathbf{r}} + \frac{m^{2}(\eta^{2m}+1)}{\eta^{2m}-1}\sin m\theta \,\hat{\boldsymbol{\theta}}\bigg]
\qquad \mathrm{at}\,\,\, r=\eta.  \label{eq:order1dynBC}
\end{split}
\end{equation}}Finally, the linearized kinematic boundary condition (\ref{eq:kinematicBC}) reads
\begin{equation}
\frac{\partial \zeta_1}{\partial t} = \hat{\mathbf{r}}\cdot\mathbf{u}_{1}  \qquad \mathrm{at}\,\,\, r=\eta. \label{eq:order1kinBC}
\end{equation}
\textcolor{black}{In this linearized problem, Eq.~(\ref{eq:order1kinBC}) is an amplitude evolution equation on the inner circle, driven by the solution of the two boundary value problems inside the inner circle and the outer annulus. This reduction allows the use of the Lorentz reciprocal theorem in the circular geometry as we describe next.}

\subsection{Lorentz reciprocal theorem \label{sec:Lorentz}}

We apply the Lorentz reciprocal theorem \cite{masoud2019reciprocal} to the linear problem of Eqs.~(\ref{eq:order1flow})--(\ref{eq:order1kinBC}) in the annular geometry with a circular boundary. To this end, we introduce an auxiliary flow problem in the same annular geometry, in which both fluids are passive and the flow is driven by a prescribed stress jump at the interface.  In this problem, both fluids are Newtonian with viscosity ratio $\lambda=\overline{\mu}/\mu$ and the interface between them is circular and centered at the origin ($\zeta=0$). We use uppercase fonts ($\mathbf{U}$, $P$, $\boldsymbol{\Sigma}^h$,...) for all variables associated with the auxiliary problem. The flow in both fluid regions is governed by the homogeneous Stokes equations,
\begin{align}
&\nabla \cdot {\mathbf{U}}=0, \qquad \nabla \cdot \boldsymbol{\Sigma}^h = \mathbf{0}, &  \mathrm{for} \,\,\, r> \eta, \label{eq:recipflow} \\
&\nabla \cdot \overline{\mathbf{U}}=0, \qquad \nabla \cdot \overline{\boldsymbol{\Sigma}}^h = \mathbf{0}, & \mathrm{for} \,\,\, r< \eta. \label{eq:recipflow2}
\end{align}
Here, $\boldsymbol{\Sigma}^h=-P\mathbf{I} + 2\,\mathbf{E}$ and $\overline{\boldsymbol{\Sigma}}^h=-\overline{P}\mathbf{I} + 2\lambda\,\overline{\mathbf{E}}$ are the Newtonian stress tensors, where $\mathbf{E}$ and $\overline{\mathbf{E}}$ are the \textcolor{black}{respective} rate-of-strain tensors. The velocity satisfies the no-slip condition at the outer boundary: $\mathbf{U}(r=1)=\mathbf{0}$, and is continuous at the interface: $\mathbf{U}(r=\eta)=\overline{\mathbf{U}}(r=\eta)$. The flow is driven by a prescribed jump in normal tractions at the interface:
\begin{equation}
\hat{\mathbf{r}}\cdot \big(\boldsymbol{\Sigma}^h-\overline{\boldsymbol{\Sigma}}^h  \big)= \Sigma_0 \cos m\theta \,\hat{\mathbf{r}},  \label{eq:stressjump}
\end{equation}
where $\Sigma_0$ is an arbitrary constant. In two dimensions, this Newtonian \textcolor{black}{fluid} problem is readily solved in terms of two streamfunctions $\psi(r,\theta)$ and $\overline{\psi}(r,\theta)$ such that 
\begin{equation}
\psi(r,\theta) = \Sigma_0 G_{m}(r) \sin m \theta, \qquad
U_r = \frac{1}{r}\frac{\partial \psi}{\partial \theta} =  \Sigma_0 \frac{G_{m}(r)}{r}m\cos m\theta, \qquad U_\theta = -\frac{\partial \psi}{\partial r}= -  \Sigma_0 G_{m}'(r) \sin m \theta,  \label{eq:streamf}
\end{equation}
with similar expressions for \textcolor{black}{the streamfunction and its derivatives} inside the drop. The two dimensionless functions $G_{m}(r)$ and $\overline{G}_{m}(r)$ are obtained analytically by solving the Stokes equations; details are presented in the Appendix. 

Using Eqs.~(\ref{eq:order1flow}) and (\ref{eq:recipflow}) in the outer \textcolor{black}{nematic} fluid region $\Omega_{0}^+=\{(r,\theta)\,|\,r>\eta\}$ and relying on the symmetry of the stress tensors, it is straightforward to derive the two relations
\begin{align}
\nabla\cdot \left[( \boldsymbol{\sigma}^h_1 +\boldsymbol{\sigma}^a_1+\boldsymbol{\sigma}^e_1)\cdot \mathbf{U}\right]&=( \boldsymbol{\sigma}^h_1 +\boldsymbol{\sigma}^a_1+\boldsymbol{\sigma}^e_1): \mathbf{E} ,  \label{eq:recip1} \\
\nabla\cdot \left[\boldsymbol{\Sigma}^h\cdot \mathbf{u}_1\right]&=\boldsymbol{\Sigma}^h: \mathbf{e}_1.  \label{eq:recip2}
\end{align}
Subtracting Eq.~(\ref{eq:recip2}) from Eq.~(\ref{eq:recip1}), integrating over the outer region $\Omega_{0}^+$, and applying the divergence theorem then yields
\begin{equation}
\int_{\mathcal{C}_{0}} \hat{\mathbf{r}}\cdot  \left[  \boldsymbol{\Sigma}^h\cdot \mathbf{u}_1 - ( \boldsymbol{\sigma}^h_1 +\boldsymbol{\sigma}^a_1+\boldsymbol{\sigma}^e_1)\cdot \mathbf{U}  \right]\mathrm{d}\ell = \int_{\Omega_{0}^+} ( \boldsymbol{\sigma}^a_1+\boldsymbol{\sigma}^e_1): \mathbf{E} \, \mathrm{d}S,  \label{eq:recipint1}
\end{equation}
where $\mathcal{C}_{0}$ denotes the circular interface at $r=\eta$. Note that the boundary term at $r=1$ vanished due to the no-slip condition, and that we used that $\boldsymbol{\sigma}^h_1: \mathbf{E} = \boldsymbol{\Sigma}^h: \mathbf{e}_1$ to simplify the right-hand side. Following an identical process in the inner region $\Omega_{0}^-=\{(r,\theta)\,|\,r<\eta\}$, we obtain,
\begin{equation}
\int_{\mathcal{C}_{0}} \hat{\mathbf{r}}\cdot  \left[ \overline{\boldsymbol{\Sigma}}^h\cdot \mathbf{u}_1 -   \overline{\boldsymbol{\sigma}}^h_1\cdot \mathbf{U} \right]\mathrm{d}\ell = 0,   \label{eq:recipint2}
\end{equation}
which does not involve any bulk integral since active and elastic stresses are absent inside $\Omega_{0}^-$. Combining Eqs.~(\ref{eq:recipint1}) and (\ref{eq:recipint2}), we obtain 
\begin{equation}
\int_{\mathcal{C}_{0}} \hat{\mathbf{r}}\cdot  \left[ \big( \boldsymbol{\Sigma}^h-  \overline{\boldsymbol{\Sigma}}^h\big)\cdot \mathbf{u}_1 - \big( \boldsymbol{\sigma}^h_1 +\boldsymbol{\sigma}^e_1+\boldsymbol{\sigma}^a_1- \overline{\boldsymbol{\sigma}}^h_1 \big)\cdot \mathbf{U}  \right]\mathrm{d}\ell = \int_{\Omega_{0}^+} \big( \boldsymbol{\sigma}^e_1+\boldsymbol{\sigma}^a_1\big): \mathbf{E} \, \mathrm{d}S,   \label{eq:reciptheorem}
\end{equation}
which is the central statement of the reciprocal theorem. The first contribution to the contour integral on the left-hand side can be simplified using the linearized kinematic boundary condition (\ref{eq:order1kinBC}) along with the auxiliary stress balance (\ref{eq:stressjump}), yielding
\begin{equation}
\int_{\mathcal{C}_{0}} \hat{\mathbf{r}}\cdot  \big( \boldsymbol{\Sigma}^h-  \overline{\boldsymbol{\Sigma}}^h\big)\cdot \mathbf{u}_1 \, \mathrm{d}\ell = \int_0^{2\pi} \Sigma_0 \cos m \theta \,\left(\hat{\mathbf{r}}\cdot \mathbf{u}_1\right)\, \eta\,  \mathrm{d}\theta = \pi \eta \Sigma_0 \frac{\mathrm{d}\widetilde{\zeta}_1}{\mathrm{d}t}.   \label{eq:surf1}
\end{equation}
Similarly, the second term in the contour integral can be simplified using the linearized dynamic boundary condition (\ref{eq:order1dynBC}) along with Eq.~(\ref{eq:streamf}) for the auxiliary velocity:
\textcolor{black}{\begin{align}
\int_{\mathcal{C}_{0}} \hat{\mathbf{r}}\cdot  \big( \boldsymbol{\sigma}^h_1 +\boldsymbol{\sigma}^e_1+\boldsymbol{\sigma}^a_1- \overline{\boldsymbol{\sigma}}^h_1 \big)\cdot \mathbf{U} \,\mathrm{d}\ell &= \int_0^{2\pi} \bigg\{\frac{\kappa_1 U_r}{\mathrm{Ca}}+\frac{\widetilde{\zeta}_1 S}{\eta}\big(U_r \cos m\theta + U_\theta m \sin m\theta \big)  \\ & \,\,\,\,\,\,- \widetilde{\zeta_1 }\frac{\xi^{2}}{\eta^{3}}\bigg[\bigg(\frac{m^{3}(\eta^{2m}+1)}{\eta^{2m}-1} + m^{2}-1\bigg) U_{r}\cos m\theta   + \frac{m^{2}(\eta^{2m}+1)}{\eta^{2m}-1} U_{\theta}\sin m\theta \bigg]  \bigg\}\eta\,\mathrm{d}\theta \nonumber \\
\begin{split}
&=m\pi \Sigma_0 \widetilde{\zeta}_1\bigg\{\left[\frac{m^2-1}{\mathrm{Ca}}\frac{G_{m}(\eta)}{\eta^2}-S  \left(G_{m}'(\eta)-\frac{G_{m}(\eta)}{\eta}\right)\right]  \\ & \,\,\,\,\,\,+ \frac{\xi^{2}}{\eta^{2}}\bigg[\frac{m(\eta^{2m}+1)}{\eta^{2m}-1}  G'_{m}(\eta)-\bigg(\frac{m^{3}(\eta^{2m}+1)}{\eta^{2m}-1} + m^{2}-1\bigg)\frac{ G_{m}(\eta)}{\eta}   \bigg]\bigg\} . 
\end{split}\label{eq:surf2}
\end{align}}\textcolor{black}{Finally, the surface integral on the right-hand side of Eq.~(\ref{eq:reciptheorem}) can be evaluated using Eqs.~(\ref{eq:sigmaalin}) and (\ref{eq:sigmaelin}) for the linearized active and elastic stresses:
\begin{align}
\int_{\Omega_{0}^+} \big( \boldsymbol{\sigma}^e_1+\boldsymbol{\sigma}^a_1\big): \mathbf{E} \, \mathrm{d}S &= \int_0^{2\pi} \int_\eta^1\bigg[ 2 S \beta_1 E_{r\theta} -\frac{2\xi^2}{r^2}\big(\partial_\theta \beta_1  E_{\theta\theta} + r\partial_{r}\beta_1 E_{r\theta}\big)   \bigg]   r\mathrm{d}r \mathrm{d}\theta  \\
\begin{split}
&= -\pi \Sigma_0  \widetilde{\zeta}_1  \frac{m \eta^{m-1}}{\eta^{2m}-1}\left\{ S\int_\eta^1 \big(r^m - r^{-m}  \big)\bigg[r G_{m}''(r)-G_{m}'(r)+m^2 \frac{G_{m}(r)}{r}  \bigg]\mathrm{d}r \right.\\
&\,\,\,\,\,\, -2m^{2}\xi^2\int_\eta^1 \big(r^{m-2} - r^{-m-2}  \big)\bigg[G_{m}'(r)-\frac{G_{m}(r)}{r} \bigg] \mathrm{d}r  \\
&\,\,\,\,\,\, - m\xi^2\left.\int_\eta^1\big(r^{m-2} + r^{-m-2}  \big)\bigg[r G_{m}''(r)-G_{m}'(r)+m^2 \frac{G_{m}(r)}{r}  \bigg] \mathrm{d}r \right\}.
\end{split}  \label{eq:volume}
\end{align}}
\subsection{Dispersion relation and growth rates}

Upon inserting Eqs.~(\ref{eq:surf1}), (\ref{eq:surf2}) and (\ref{eq:volume}) into the reciprocal relation (\ref{eq:reciptheorem}), the arbitrary stress jump magnitude $\Sigma_0$ can be eliminated, and we arrive at a first-order ODE for the interfacial perturbation amplitude,
\begin{equation}
\frac{\mathrm{d}\widetilde{\zeta}_1}{\mathrm{d}t}=\Lambda_m \widetilde{\zeta}_1.
\end{equation}
The constant growth rate $\Lambda_m$ for normal mode $m$ includes contributions from capillary, active and elastic stresses:
\begin{equation}
\Lambda_m = \frac{1}{\mathrm{Ca}} \Xi^c_m (\eta,\lambda) + S\,  \Xi^a_m (\eta,\lambda) + \xi^2\,  \Xi^e_m (\eta,\lambda),  \label{eq:Lambda}
\end{equation}
respectively given by
\begin{align}
 \Xi^c_m (\eta,\lambda) &= m(m^2-1) \frac{G_{m}(\eta)}{\eta^3},\\
 \Xi^a_m (\eta,\lambda) &= -m\frac{\mathrm{d}}{\mathrm{d}\eta}\bigg[\frac{G_{m}(\eta)}{\eta}\bigg]- \frac{m \eta^{m-2}}{\eta^{2m}-1}\int_\eta^1 \big(r^m - r^{-m}  \big)\bigg[r G_{m}''(r)-G_{m}'(r)+m^2 \frac{G_{m}(r)}{r}  \bigg]\mathrm{d}r, \label{eq:Xia} \\ 
  \begin{split}
 \Xi^e_m (\eta,\lambda) & =  \textcolor{black}{-\frac{m}{\eta^{3}}\bigg[\frac{m(\eta^{2m}+1)}{\eta^{2m}-1}  G'_{m}(\eta)-\bigg(\frac{m^{3}(\eta^{2m}+1)}{\eta^{2m}-1} + m^{2}-1\bigg)\frac{ G_{m}(\eta)}{\eta}   \bigg]}  \\ & \,\,\,\,\,\,\,\textcolor{black}{+ \frac{m \eta^{m-2}}{\eta^{2m}-1}\bigg\{2m^{2}\int_\eta^1 \big(r^{m-2}  - r^{-m-2}  \big)\bigg[G_{m}'(r)-\frac{G_{m}(r)}{r} \bigg] \mathrm{d}r}    \\ & \,\,\,\,\,\,\,\textcolor{black}{+ m\int_\eta^1\big(r^{m-2} + r^{-m-2}  \big)\bigg[r G_{m}''(r)-G_{m}'(r)+m^2 \frac{G_{m}(r)}{r}  \bigg] \mathrm{d}r \bigg\}}.  \end{split}  \label{eq:Xie}
\end{align}
These various growth rates depend on the function $G_{m}(r)$ appearing in the solution of the Newtonian auxiliary problem, whose form differs between the translation ($m=1$) and deformation ($m\ge 2$) modes of the drop. We analyze and discuss these two cases separately. 

\subsubsection{Translation mode $(m=1)$}

The first normal mode $m=1$ captures translation of the drop and is the only mode with a non-zero displacement of the drop center of mass.\ At linear order, it does not affect the curvature of the drop, and therefore the growth is unaffected by surface tension: $\Xi_1^c(\eta,\lambda)=0$.\ The function $G_{m}(r)$ for $m=1$ is given in Eq.~(\ref{eq:Gtrans}).\ Inserting this expression into Eqs.~(\ref{eq:Xia}) and (\ref{eq:Xie}) and simplifying provides the active and elastic growth rates,
\begin{align}
\Xi^a_1(\eta,\lambda)&=-2 \eta\, \overline{A}_1 - \frac{1-\eta^2}{\eta}A_1-2 \left[\frac{1}{\eta^3}+\frac{2\ln \eta}{\eta(1-\eta^2)}  \right]D_1 , \\
\Xi^e_1(\eta,\lambda)&=\textcolor{black}{-2\bigg(\frac{\eta^{2}+1}{\eta^{2} -1}\bigg)\,\frac{\overline{A}_1}{\eta} -\frac{4}{\eta}A_{1} + \left[\frac{1}{\eta^3}+\frac{2 \ln \eta}{\eta(1-\eta^2)}\right]B_1-2\frac{1+\eta^2}{\eta^5}D_1},
\end{align}
where the coefficients $A_1$, $B_1$, $D_1$ and $\overline{A}_1$ are dimensionless functions of $\eta$ and $\lambda$ given by Eqs.~(\ref{eq:A1})--(\ref{eq:A1bar}) of the Appendix. 

\begin{figure}[t]
\centering
\includegraphics[width=0.98\textwidth]{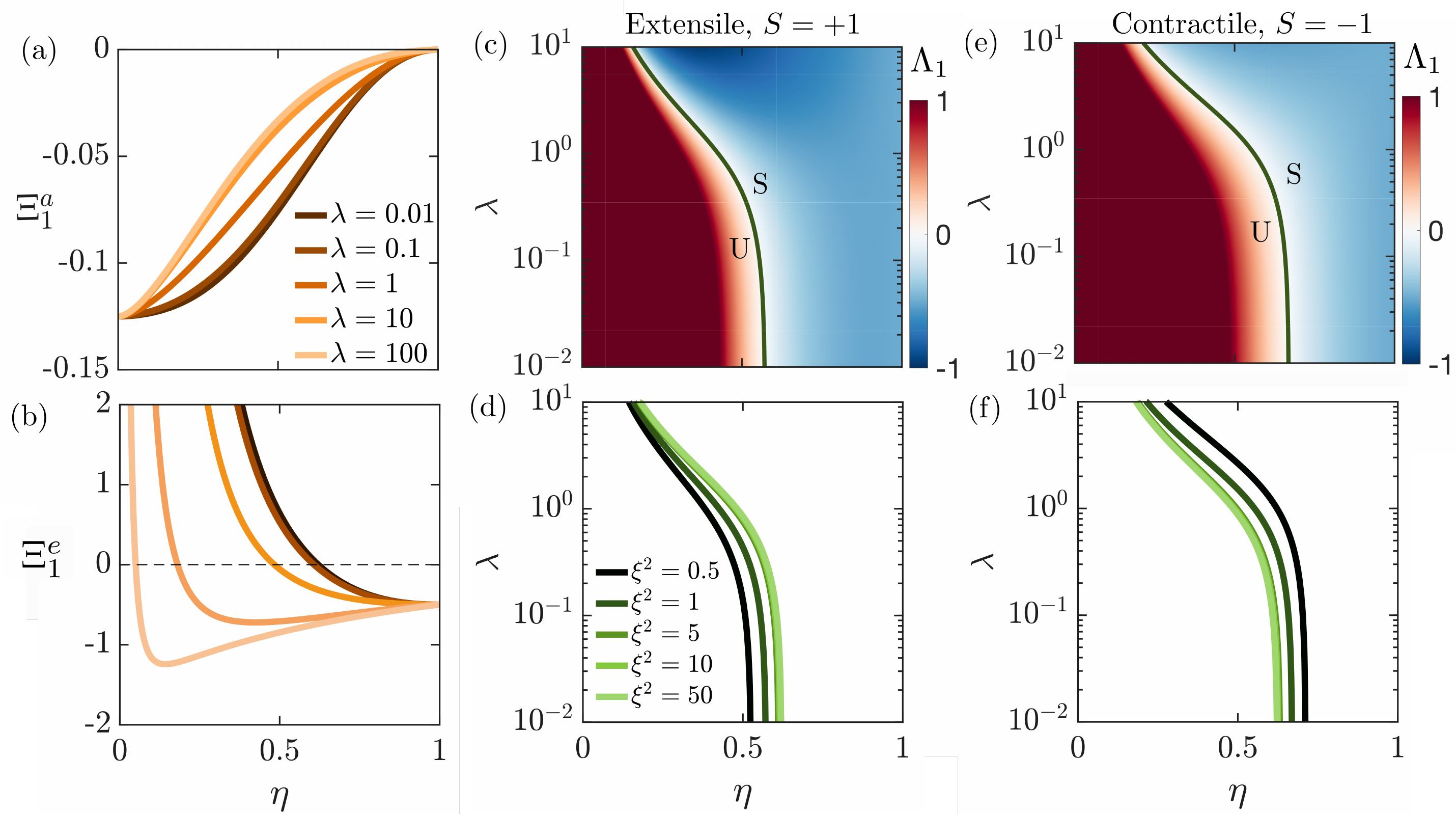}\vspace{-0.2cm}
\caption{(a) Active contribution $\Xi^a_1$ and (b) elastic contribution $\Xi^e_1$ to the growth rate $\Lambda_1$ for the translational mode ($m=1$), plotted as functions of $\eta$ for different viscosity ratios $\lambda$. (c) Color map of the total growth rate $\Lambda_1$ in the $(\eta,\lambda)$ parameter space for an extensile system ($S=+1$) with $\xi^2=1$. Linearly stable (S) and unstable (U) regions are shown in blue and red, respectively, and the green curve shows the marginal stability curve. (d) Effect of $\xi^2$ on the marginal stability curve in (c). Panels (e) and (f) show the same as (c) and (d) for a contractile system ($S=-1$).}
\label{fig:translationgrowthrates} \vspace{-0.3cm}
\end{figure}


The two functions $\Xi^a_1$ and $\Xi^e_1$ are plotted vs $\eta$ for different viscosity ratios in Fig.~\ref{fig:translationgrowthrates}(a,b). As seen in Fig.~\ref{fig:translationgrowthrates}(a), $\Xi^a_1(\eta,\lambda)<0$ for all values of $\eta$ and $\lambda$, indicating that the translational mode is linearly stable under active stresses in extensile systems, but unstable in contractile systems (recall that $\Xi^a_1$ gets multiplied by $S$ in Eq.~(\ref{eq:Lambda}) for the actual growth rate). $\Xi^a_1$ only shows a weak dependence on viscosity ratio. The two asymptotes for a rigid particle ($\lambda\to\infty$) and an inviscid bubble ($\lambda\to0$) are given by 
\begin{equation}
\lim_{\lambda\to\infty} \Xi^a_1=-\frac{1}{8}\left[1+\frac{4\eta^2\ln\eta}{1-\eta^4}\right], \qquad \lim_{\lambda\to 0} \Xi^a_1=-\frac{1}{8}\left[1-\frac{\eta^2(1-\eta^2)^2 - 4 \eta^4 \ln \eta}{(1-\eta^2)(1+\eta^4)}    \right]. 
\end{equation}
The elastic contribution shows a more complex dependence on both $\eta$ and $\lambda$ in Fig.~\ref{fig:translationgrowthrates}(b). \textcolor{black}{Larger drops are always stable under elastic stresses, especially at high viscosity ratios.  Regardless of $\lambda$, we find, however, that $\Xi^e_1\to +\infty$ as $\eta\to 0$, i.e., small drops are always unstable to translation.} The divergence of the growth rate in that limit is attributed to the divergence of the base-state elastic pressure at the origin when $\eta\to 0$ as seen in Eq.~(\ref{eq:basepressurenematic}). When both active and elastic stresses play a role, the translational mode can be either stable or unstable depending on parameter values. For the extensile case, this is illustrated in Fig.~\ref{fig:translationgrowthrates}(c), showing a color plot of the total growth rate $\Lambda_1$ in the $(\eta,\lambda)$ parameter space: \textcolor{black}{the system is always unstable for small drops and becomes stable as $\eta$ is increased, with the extent of the unstable region shrinking with increasing viscosity ratio.} The dependence of the marginal stability curve on $\xi^2$ is shown in Fig.~\ref{fig:translationgrowthrates}(d), where increasing the magnitude of elastic stresses is found to increase the size of the unstable region. The contractile case is analyzed similarly in Fig.~\ref{fig:translationgrowthrates}(e,f): the unstable region is found to be larger in that case, but to decrease in size with increasing $\xi^2$.

\begin{figure}[t]
\centering
\includegraphics[width=0.98\textwidth]{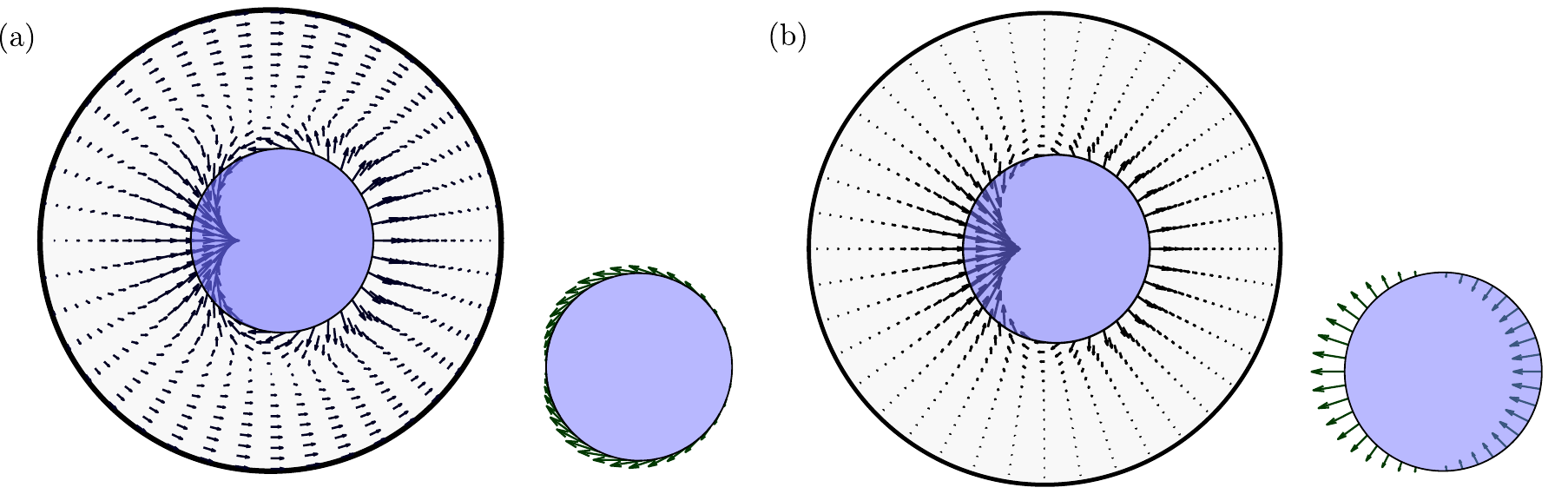}
\caption{(a) Active and (b) elastic bulk force densities and surface tractions at linear order in $\epsilon$ for the translation mode $m=1$, in the case of a drop with radius $\eta=0.45$. In each case, the left plot shows the first-order force density, $\mathbf{f}^{a,e}_1=\nabla\cdot \boldsymbol{\sigma}^{a,e}_1$, in the bulk of the nematic, while the right plot shows the first-order traction field, \textcolor{black}{$\mathbf{n}_{0}\cdot \boldsymbol{\sigma}_{1}^{a} + \mathbf{n}_{1}\cdot \boldsymbol{\sigma}_{0}^{a} $ or $\mathbf{n}_{0}\cdot \boldsymbol{\sigma}_{1}^{e} + \mathbf{n}_{1}\cdot \boldsymbol{\sigma}_{0}^{e}+\mathbf{t}_{s,1}^e $,} on the drop surface. Panel (a) shows the active force and traction for an extensile system with $S=+1$.}
\label{fig:forcefieldsbulkandsurface}
\end{figure}

Some partial intuition into the trends observed in Fig.~\ref{fig:translationgrowthrates} can be obtained by analyzing the active and elastic stresses exerted in the bulk of the nematic and on the interface. The active and elastic force densities exerted in the bulk are given by the divergence of the respective stress tensors, $\mathbf{f}^{a} = \nabla\cdot \boldsymbol{\sigma}^{a}$ and $\mathbf{f}^{e} = \nabla\cdot \boldsymbol{\sigma}^{e}$. At linear order and for $m=1$, their $(r,\theta)$ components are given by
\begin{equation}
f^{a}_{r1} = (\eta^2 -1)^{-1} ( 1 - r^{-2})\cos \theta, \qquad f^{a}_{\theta 1} = (\eta^2 -1)^{-1}(3-r^{-2})\sin \theta,
\end{equation}
and
\textcolor{black}{\begin{equation}
f^{e}_{r1} = (\eta^2 -1)^{-1} ( r^{-2} - 3r^{-4})\cos \theta, \qquad f^{e}_{\theta 1} = (\eta^2 -1)^{-1}(r^{-2} - r^{-4})\sin \theta .
\end{equation}}The two force fields are shown in Fig.~\ref{fig:forcefieldsbulkandsurface} for a representative case with $\eta=0.45$, for a circular drop that was displaced to the right from its equilibrium position; the active force field shown here is for an extensile system and changes sign in the contractile case. The active and elastic force bulk densities display a complex structure and are strongest near the drop surface, where they would appear to have a destabilizing effect by pushing the fluid and drop further to the right. It should be kept in mind, however, that these force fields are applied on the solvent in the nematic layer, where they drive a flow whose structure is not explicitly known as the Lorentz reciprocal theorem circumvented its calculation. First-order active and elastic tractions on the interface, \textcolor{black}{$\mathbf{n}_{0}\cdot \boldsymbol{\sigma}_{1}^{a} + \mathbf{n}_{1}\cdot \boldsymbol{\sigma}_{0}^{a} $ and $\mathbf{n}_{0}\cdot \boldsymbol{\sigma}_{1}^{e} + \mathbf{n}_{1}\cdot \boldsymbol{\sigma}_{0}^{e}+\mathbf{t}_{s,1}^e $, respectively,}  are shown in the lower-right panels in Fig.~\ref{fig:forcefieldsbulkandsurface}, where they both appear to resist translation of the drop and thus have a stabilizing effect. Here again, these traction fields do not account for the additional hydrodynamic tractions resulting from flow fields inside and outside the drop, and thus provide an incomplete description of stresses acting on the interface.\ Nevertheless, the stresses in Fig.~\ref{fig:forcefieldsbulkandsurface} suggest a subtle interplay of active, elastic and viscous effects in governing the growth rates calculated in Fig.~\ref{fig:translationgrowthrates}.

%

\subsubsection{Deformation modes $(m\ge 2)$}

Modes with $m\ge 2$ describe deformations of the drop interface and do not involve translation of the center of mass. The function $G_{m}(r)$ for $m\ge 2$ is given in Eq.~(\ref{eq:Gdef}). The corresponding capillary, active, and elastic contributions to the growth rate can be obtained as
\begin{align}
\Xi^c_m(\eta,\lambda)&=m(m^2-1)\left[ \eta^{m-3}\,\overline{A}_m+ \eta^{m-1}\,\overline{C}_m\right], \label{eq:sigmac1} \\[10pt]
\begin{split}
\Xi^a_m(\eta,\lambda)&=-m\left[(m-1)  \eta^{m-2}\,\overline{A}_m+(m+1) \eta^{m}\,\overline{C}_m  \right]  \\[5pt]
&\,\,\,\,\,\,- \frac{m^2 \eta^{m-2}}{\eta^{2m}-1}\left\{ (m-1) \left[\frac{1-\eta^{2m}}{m}+2\ln \eta\right]A_m+ (m+1)\left[\frac{1-\eta^{-2m}}{m}-2\ln \eta\right]B_m\right.    \\
&\,\,\,\,\,\, \left. + (m+1) \left[\frac{1-\eta^{2+2m}}{m+1}-1+\eta^2\right]C_m+ (m-1)  \left[\frac{1-\eta^{2-2m}}{m-1}+1-\eta^2\right]D_m\right\}, 
\end{split} \\[10pt]
\begin{split}
\textcolor{black}{\Xi^e_m(\eta,\lambda)}&=\textcolor{black}{m\left[(m-1)\bigg(\frac{m^{2}(\eta^{2m}+1)}{\eta^{2m}-1}+m+1\bigg)  \eta^{m-4}\,\overline{A}_m+\bigg(\frac{m^{3}(\eta^{2m}+1)}{\eta^{2m}-1}- 2m-1\bigg) \eta^{m-2}\,\overline{C}_m  \right]} \\[5pt]
&\textcolor{black}{\,\,\,\,\,\,+\frac{2m^2 \eta^{m-2}}{\eta^{2m}-1}\bigg[m(1 - \eta^{2m-2})A_{m}-m(1 - \eta^{-2(1+m)})B_{m} }\\ &\,\,\,\,\,\,\textcolor{black}{ +    (1+m)(1-\eta^{2m})C_{m}-(m-1)(1-\eta^{-2m})D_{m}\bigg]}, 
\end{split}
\end{align}
where the flow coefficients $A_m$, $B_m$ $C_m$, $D_m$, $\overline{A}_m$ and $\overline{C}_m$ are provided in Eqs.~(\ref{eq:Am})--(\ref{eq:Cmbar}) of the Appendix.

\begin{figure}[t]
\centering
\includegraphics[width=0.95\textwidth]{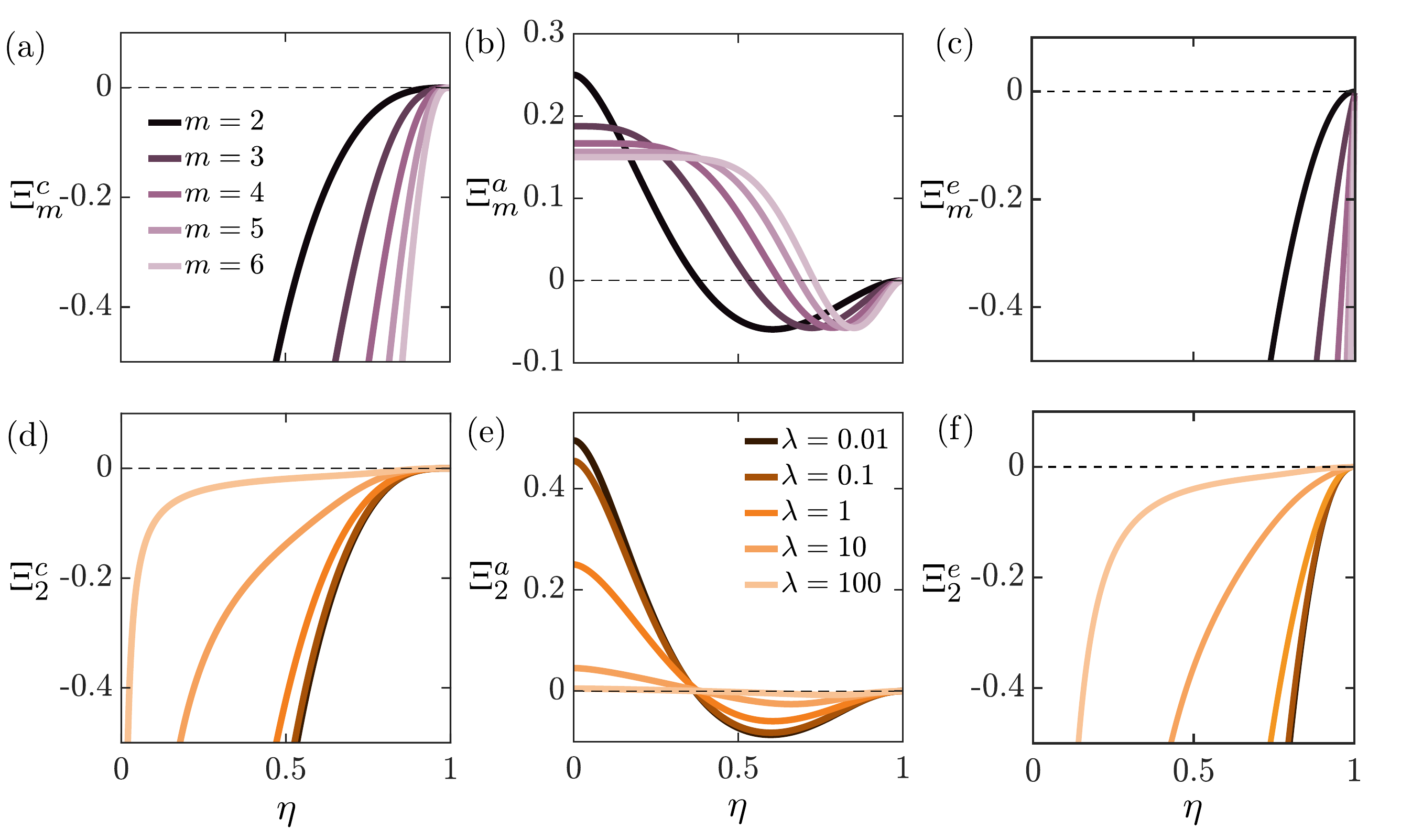} \vspace{-0.0cm}
\caption{(a) Capillary contribution $\Xi^a_m$, (b) active contribution $\Xi^a_m$, and (c) elastic contribution $\Xi^e_m$ to the growth rate $\Lambda_m$ for translational modes of increasing wavenumber $m$, plotted as functions of $\eta$ for $\lambda=1$. Panels (d), (e) and (f) show the same functions for mode $m=2$ and for varying values of the viscosity ratio $\lambda$.}
\label{fig:deformationgrowthrates}
\end{figure}

\begin{figure}[t]
\centering
\includegraphics[width=0.94\textwidth]{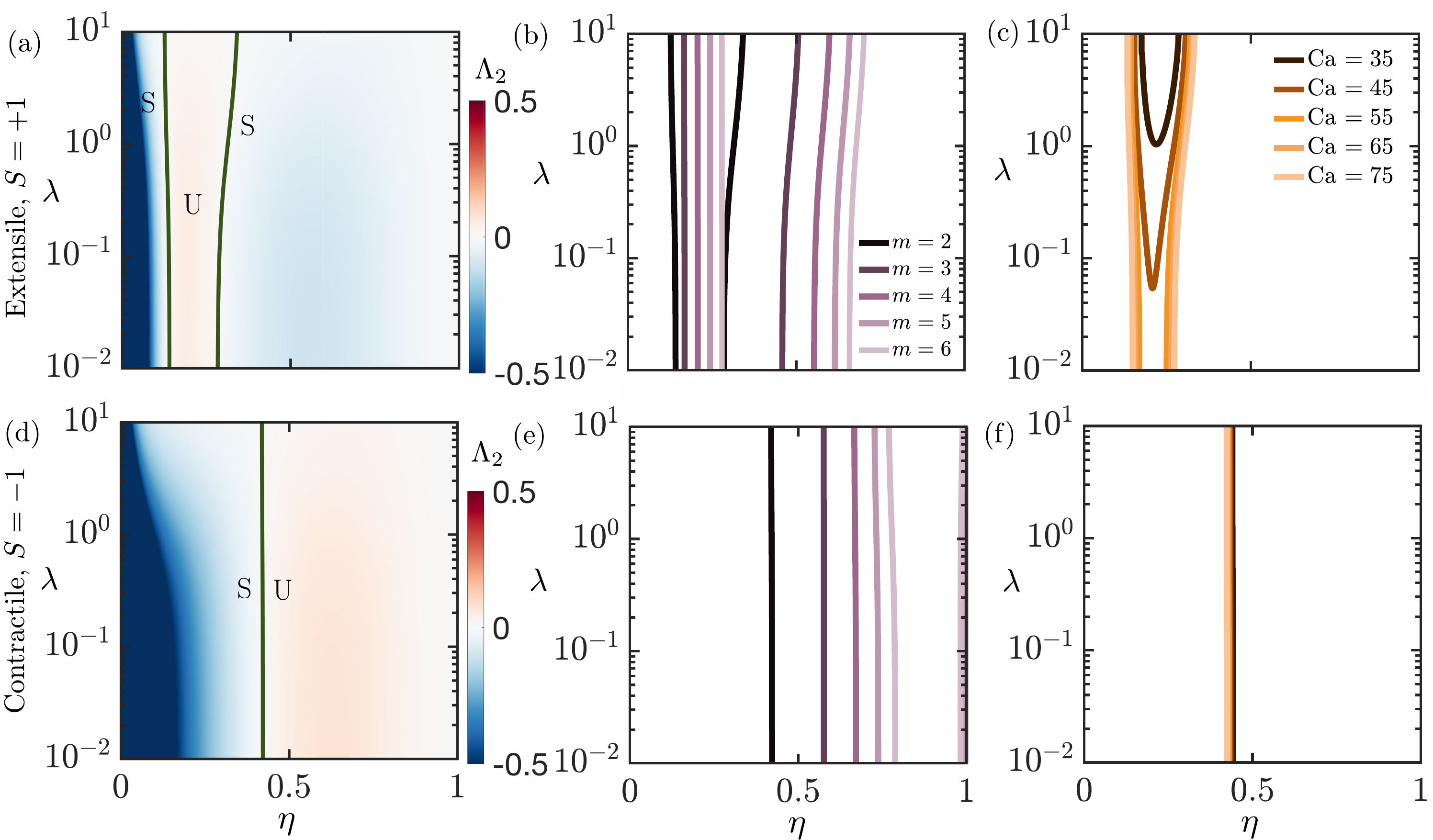} \vspace{-0.25cm}
\caption{(a) Color map of the total growth rate $\Lambda_2$ in the $(\eta,\lambda)$ parameter space for an extensile system ($S=+1$) with $\xi^2=0.005$ and Ca $=100$. Linearly stable (S) and unstable (U) regions are shown in blue and red, respectively, and the \textcolor{black}{ \textcolor{black}{green}} curve shows the marginal stability curve.  Effect of (b) $m$ and (c) Ca on the marginal stability curve in (a). Panels (d), (e), and (f) show the same as (a), (b), and (c) for a contractile system ($S=-1$).}
\label{fig:deformationgrowthrates_marginalstability}
\end{figure}

The dependence of $\Xi^c_m$, $\Xi^a_m$ and $\Xi^e_m$ on $m$, $\eta$ and $\lambda$ is illustrated in Fig.~\ref{fig:deformationgrowthrates}, where panels (a), (b) and (c) show the dependence on drop size $\eta$ for distinct modes $m$, and panels (d), (e) and (f) show the effect of viscosity ratio $\lambda$ for mode $m=2$. Capillary stresses always favor a circular drop and are therefore stabilizing, especially in the case of small drops and high wavenumbers; this damping effect of surface tension is most pronounced in low-viscosity drops. In the case of extensile systems, activity is destabilizing in the case of drops of small to intermediate sizes ($0\le \eta\lesssim 0.5$), and stabilizing for large drops, and the range of unstable sizes increases with $m$; these trends are reversed for contractile systems. In both cases, low-viscosity drops are the ones that are most affected by activity (largest magnitude of the growth rate), and the limit of high-viscosity drops ($\lambda\to \infty$) is neutrally stable. On the other hand, \textcolor{black}{nematic elasticity is always stabilizing, especially in the case of small drops.} These effects are most pronounced at low wavenumbers and low viscosity ratios. 

Figure~\ref{fig:deformationgrowthrates_marginalstability}(a) shows a color plot of the total growth rate $\Lambda_2$ corresponding to mode $m=2$ in the $(\eta,\lambda)$ parameter space, for an extensile system.\ The system is stable at lower values of the radius ratio ($0\leq\eta\lesssim 0.1$) followed by a range of unstable droplet sizes $(0.1\lesssim\eta\lesssim 0.25)$, and eventually regains stability at larger values of $\eta$.\ This behavior can be explained as follows: at lower values of $\eta$, capillary and elastic stresses dominate and are stabilizing; as $\eta$ increases, their effect weakens and the system becomes unstable as a result of active stresses; at larger values of $\eta$, active stresses become stabilizing leading to a negative growth rate. The marginal stability curve demarcating these regimes is only weakly affected by the viscosity ratio. As shown in Fig.~\ref{fig:deformationgrowthrates_marginalstability}(b), increasing the mode number $m$ tends to increase the size of the unstable region.\ 
Expectedly, decreasing the capillary number tends to stabilize the system across all drop sizes, especially at low viscosity ratios as illustrated in Fig.~\ref{fig:deformationgrowthrates_marginalstability}(c).
The contractile case is analyzed similarly in Fig.~\ref{fig:deformationgrowthrates_marginalstability}(d,e,f). In this case, alternating regions of stable and unstable growth rates are not observed; rather, a single stable region at low values of $\eta$ gives way to an unstable region in the case of larger drops.

\begin{figure}[t]
\centering
\includegraphics[width=0.98\textwidth]{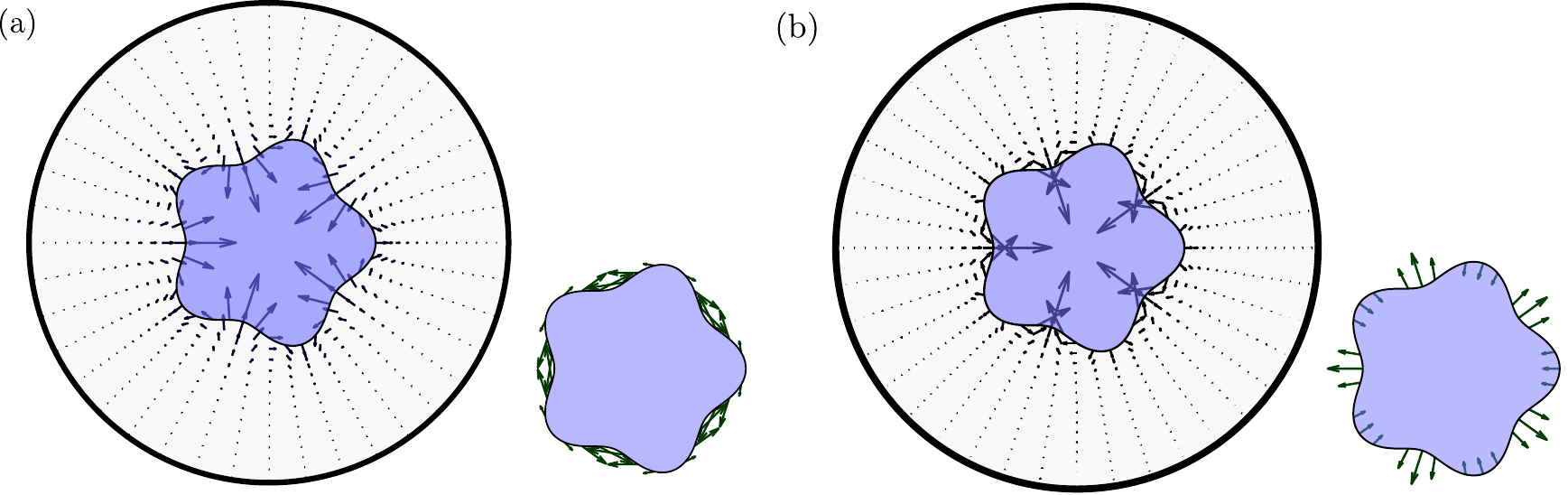}
\caption{(a) Active and (b) elastic bulk force densities and surface tractions at linear order in $\epsilon$ for the deformation mode $m=5$, in the case of a drop with radius $\eta=0.45$. In each case, the left plot shows the first-order force density, $\mathbf{f}^{a,e}_1=\nabla\cdot \boldsymbol{\sigma}^{a,e}_1$, in the bulk of the nematic, while the right plot shows the first-order traction field, \textcolor{black}{$\mathbf{n}_{0}\cdot \boldsymbol{\sigma}_{1}^{a} + \mathbf{n}_{1}\cdot \boldsymbol{\sigma}_{0}^{a} $ or $\mathbf{n}_{0}\cdot \boldsymbol{\sigma}_{1}^{e} + \mathbf{n}_{1}\cdot \boldsymbol{\sigma}_{0}^{e}+\mathbf{t}_{s,1}^e $,}  on the drop surface. Panel (a) shows the active force and traction for an extensile system with $S=+1$.}
\label{fig:forcefieldsbulkandsurfacem5}
\end{figure}

As done previously for the translational mode, we examine the bulk and interfacial active and elastic stresses emerging at linear order due to the deformation of the drop. When $m\ge 2$, the bulk active and elastic force densities have components 
\begin{equation}
f^{a}_{rm} = \frac{m^{2}\eta^{m-1}}{\eta^{2m}-1}\bigg(r^{m-1} - \frac{1}{r^{m+1}}\bigg) \cos m\theta, \qquad
f^{a}_{\theta m} = \frac{m\eta^{m-1}}{\eta^{2m}-1}\bigg( (m+2)r^{m-1} + \frac{m-2}{r^{m+1}}\bigg) \sin m\theta,
\end{equation}
and
\textcolor{black}{\begin{equation}
f^{e}_{rm} = \frac{m^{2}\eta^{m-1}}{\eta^{2m}-1}\bigg((2-m)r^{m-3} - \frac{m+2}{r^{m+3}}\bigg) \cos m\theta, \qquad
f^{e}_{\theta m} = \frac{m^{3}\eta^{m-1}}{\eta^{2m}-1}\bigg(r^{m-3} - \frac{1}{r^{m+3}}\bigg)\sin m\theta .
\end{equation}}They are plotted in Fig.~\ref{fig:forcefieldsbulkandsurfacem5} for mode $m=5$ and $\eta=0.45$, along with the first-order surface tractions acting on the interface. The fields shown in the figure suggest a destabilizing effect of bulk stresses and a stabilizing effect of surface tractions, both elastic and active (in the extensile case). The interplay of these forces, together with hydrodynamic stresses whose distribution is not explicitly known, gives rise to the trends found in Figs.~\ref{fig:deformationgrowthrates} and \ref{fig:deformationgrowthrates_marginalstability}.

\section{The inverted problem: Active nematic drop within a passive layer}\label{section4}

In this section, we extend our theory to analyze the inverse scenario, namely, the stability of an active nematic droplet surrounded by a passive viscous layer. Much of the theoretical formulation presented above remains unchanged, with the difference that active and elastic stresses are now exerted inside the drop ($r<\eta R+\zeta$). As a result, we do not present the full theory again, but instead only highlight places where differences arise.


\subsection{Governing equations, boundary conditions and non-dimensionalization}

Governing equations (\ref{eq:nematodynamics1})--(\ref{eq:stokes2}) still apply, the only difference being that the active and elastic stresses of Eqs.~(\ref{eq:sigmaa}) and (\ref{eq:sigmae}) are now applied inside the drop. As a result, they are decorated by an overbar and enter the momentum equation (\ref{eq:stokes2}) instead of (\ref{eq:stokes1}). The kinematic boundary condition (\ref{eq:kinematicBC}) is unchanged, while the dynamic boundary condition, in dimensionless form, becomes: 
\begin{equation}
\mathbf{n}\cdot(\boldsymbol{\sigma}^h-\overline{\boldsymbol{\sigma}}^a-\overline{\boldsymbol{\sigma}}^e-\overline{\boldsymbol{\sigma}}^h)\textcolor{black}{\,+\,\overline{\mathbf{t}}_s^e}=\frac{1}{\mathrm{Ca}} \kappa  \mathbf{n} .   \label{eq:dynBC_invertedproblem}
\end{equation}
\textcolor{black}{Given our choice of the normal as pointing outward, the elastic traction exerted by the nematic drop on the interface is now given by 
\begin{equation}
    \overline{\mathbf{t}}_s^e=-\xi^{2} \frac{\partial }{\partial s}\bigg(\frac{\partial \beta}{\partial n}\mathbf{n}\bigg),
\end{equation}
where arclength is still defined in the counterclockwise direction. }


\subsection{Linear stability analysis}

\subsubsection{Base state}

In the base state, the nematic field lines inside the drop are also circles centered at the origin, with $\beta_0$ given by Eq.~(\ref{eq:beta0}). A topological defect of charge $+1$ is present at the origin. The base state active and elastic stresses are still given by Eq.~(\ref{eq:basestress}). They induce a pressure field inside the drop that diverges at the location of the defect, while the pressure outside is uniform:
\begin{equation}
\overline{p}_0(r) = S\ln r - \frac{\xi^{2}}{2r^{2}}, \qquad \textcolor{black}{p_0 = S\ln \eta - \frac{\xi^{2}}{2\eta^{2}} - \frac{1}{\eta \mathrm{Ca}}}. \label{eq:basepressurenematic_invertedproblem}
\end{equation}

\subsubsection{Domain perturbation and linearized problem}

The shape of the drop is once again deformed according to Eq.~(\ref{eq:shapedef}), which results in the boundary condition of Eq.~(\ref{eq:beta1bc}) for the perturbation angle $\beta_1$ at the drop surface. A solution of $\nabla^2\beta_1=0$ inside the drop that satisfies this condition while remaining bounded at $r=0$ can be obtained as
\textcolor{black}{\begin{equation}
\beta_1(r,\theta) = \widetilde{\zeta}_1 m\eta^{-m-1} r^{m}\sin m\theta. 
\end{equation}}
Equations~(\ref{eq:sigmaalin}) and (\ref{eq:sigmaelin}) are still valid inside the drop. At linear order, the dynamic boundary condition (\ref{eq:dynBC_invertedproblem}) yields
\begin{equation}
\begin{split}
\hat{\mathbf{r}}\cdot \big(\boldsymbol{\sigma}^h_1-\overline{\boldsymbol{\sigma}}^a_1 - \overline{\boldsymbol{\sigma}}^e_1- \overline{\boldsymbol{\sigma}}^h_1 \big) &=\frac{\kappa_1 \hat{\mathbf{r}}}{\mathrm{Ca}}\textcolor{black}{-\frac{\widetilde{\zeta_1}}{\eta}S\big(\cos m\theta \,\hat{\mathbf{r}}+m\sin m\theta\,\hat{\boldsymbol{\theta}} \big)} \\ & \textcolor{black}{\,\,\,\,\,\,+\widetilde{\zeta_1 }\frac{\xi^{2}}{\eta^{3}}\Big[\big(m^{3} + m^{2}-1\big)\cos m\theta \,\hat{\mathbf{r}} + m^{2} \sin m\theta \,\hat{\boldsymbol{\theta}}\Big] 
\qquad \mathrm{at}\,\,\, r=\eta.}
\end{split}
\label{eq:order1dynBC_invertedproblem}
\end{equation}

\subsubsection{Lorentz reciprocal theorem}

Application of the Lorentz reciprocal theorem follows the same steps as discussed in Sec.~\ref{sec:Lorentz} and uses the same auxiliary Newtonian flow problem. Once again, the main difference with Sec.~\ref{sec:Lorentz} is that active and elastic stresses now appear inside the drop. As a result, the analog of Eq.~(\ref{eq:recipint1}) has no bulk contribution and reads
\begin{equation}
\int_{\mathcal{C}_{0}} \hat{\mathbf{r}}\cdot  \left[  \boldsymbol{\Sigma}^h\cdot \mathbf{u}_1 -  \boldsymbol{\sigma}^h_1 \cdot \mathbf{U}  \right]\mathrm{d}\ell = 0 .  \label{eq:recipint1_invertedproblem}
\end{equation}
Conversely, the analog of Eq.~(\ref{eq:recipint2}) does include  a surface integral: 
\begin{equation}
\int_{\mathcal{C}_{0}} \hat{\mathbf{r}}\cdot  \left[ \overline{\boldsymbol{\Sigma}}^h\cdot \mathbf{u}_1 -   (\overline{\boldsymbol{\sigma}}^h_1 +\overline{\boldsymbol{\sigma}}^a_1+\overline{\boldsymbol{\sigma}}^e_1)\cdot \mathbf{U}\right]\mathrm{d}\ell = \int_{\Omega_{0}^-} ( \overline{\boldsymbol{\sigma}}^a_1+\overline{\boldsymbol{\sigma}}^e_1): \overline{\mathbf{E}} \, \mathrm{d}S,   \label{eq:recipint2_invertedproblem}
\end{equation}
where $\Omega_{0}^-=\{(r,\theta)\,|\,r<\eta\}$ denotes the interior of the drop in two dimensions. Combining Eqs.~(\ref{eq:recipint1_invertedproblem}) and (\ref{eq:recipint2_invertedproblem}) yields the main result 
\begin{equation}
\int_{\mathcal{C}_{0}} \hat{\mathbf{r}}\cdot  \left[ \big( \boldsymbol{\Sigma}^h-  \overline{\boldsymbol{\Sigma}}^h\big)\cdot \mathbf{u}_1 - \big( \boldsymbol{\sigma}^h_1 -\overline{\boldsymbol{\sigma}}^e_1-\overline{\boldsymbol{\sigma}}^a_1- \overline{\boldsymbol{\sigma}}^h_1 \big)\cdot \mathbf{U}  \right]\mathrm{d}\ell = -\int_{\Omega_{0}^-} \big( \overline{\boldsymbol{\sigma}}^e_1+\overline{\boldsymbol{\sigma}}^a_1\big): \overline{\mathbf{E}} \, \mathrm{d}S.   \label{eq:reciptheorem_invertedproblem}
\end{equation}

The first term in the contour integral can be simplified as in Eq.~(\ref{eq:surf1}).\ Using the linearized dynamic boundary condition (\ref{eq:order1dynBC_invertedproblem}), the second term simplifies to
\begin{align}
\begin{split}
\int_{\mathcal{C}_{0}} \hat{\mathbf{r}}\cdot  \big( \boldsymbol{\sigma}^h_1 -\overline{\boldsymbol{\sigma}}^e_1-\overline{\boldsymbol{\sigma}}^a_1- \overline{\boldsymbol{\sigma}}^h_1 \big)\cdot \mathbf{U} \,\mathrm{d}\ell &=m\pi \Sigma_0 \widetilde{\zeta}_1\bigg\{\frac{m^2-1}{\mathrm{Ca}}\frac{\overline{G}_{m}(\eta)}{\eta^2}+S \left(\overline{G}_{m}'(\eta)-\frac{\overline{G}_{m}(\eta)}{\eta}\right) \\ &\textcolor{black}{\,\,\,\,\,\,-\frac{\xi^{2}}{\eta^{2}}\bigg[m\overline{G}_{m}'(\eta)-(m^{3}+m^{2} -1)\frac{\overline{G}_{m}(\eta)}{\eta}    \bigg]\bigg\}}.
\end{split}
\label{eq:surf2_invertedproblem}
\end{align}


Finally, the surface integral on the right-hand side of Eq.~(\ref{eq:reciptheorem_invertedproblem}) can be evaluated as
\begin{align}
\begin{split}
\int_{\Omega_{0}^-} \big( \overline{\boldsymbol{\sigma}}^e_1+\overline{\boldsymbol{\sigma}}^a_1\big): \overline{\mathbf{E}} \, \mathrm{d}S &= -\pi \Sigma_0  \widetilde{\zeta}_1  m \textcolor{black}{\eta^{-m-1}} \left\{ S\int_0^\eta r^m \bigg[r \overline{G}_{m}''(r)-\overline{G}_{m}'(r)+m^2 \frac{\overline{G}_{m}(r)}{r}  \bigg]\mathrm{d}r \right.\\
&\,\,\,\,\,\,\textcolor{black}{ - \xi^2\int_0^\eta r^{m-2}\bigg[m r \overline{G}_{m}''(r)  +(2m^2-m)\overline{G}_{m}'(r) +(m^3-2m^2) \frac{\overline{G}_{m}(r)}{r}  \bigg] \mathrm{d}r  \bigg\}}.
\end{split}  \label{eq:volume_invertedproblem}
\end{align}



\subsection{Dispersion relation and growth rates}

Inserting Eqs.~(\ref{eq:surf1}), (\ref{eq:surf2_invertedproblem}) and (\ref{eq:volume_invertedproblem}) into Eq.~(\ref{eq:reciptheorem_invertedproblem}) provides an expression for the growth rate in the form on Eq.~(\ref{eq:Lambda}), with contributions from capillary, active and elastic stresses that we discuss next. 

\begin{figure}[t]
\centering
\includegraphics[width=1.0\textwidth]{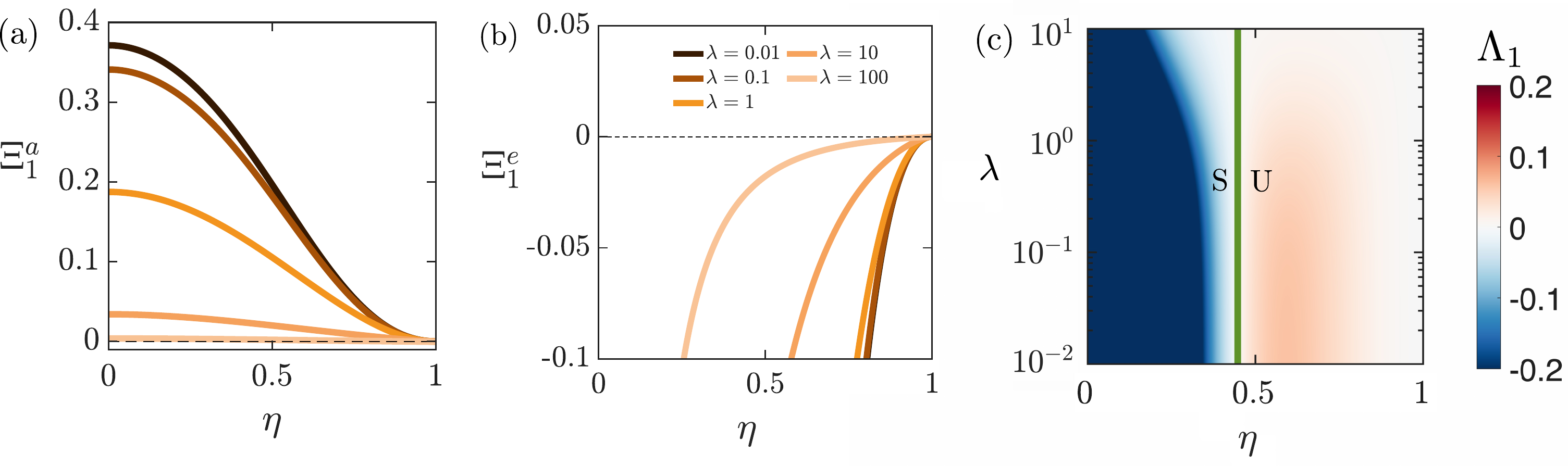}
\caption{(a) Active contribution $\Xi^a_1$ and (b) elastic contribution $\Xi^e_1$ to the growth rate $\Lambda_1$ for the translational mode ($m=1$), plotted as functions of $\eta$ for different viscosity ratios $\lambda$. (c) Color map of the total growth rate $\Lambda_1$ in the $(\eta,\lambda)$ parameter space for an extensile system ($S=+1$) with \textcolor{black}{$\xi^2=0.1$}. Linearly stable (S) and unstable (U) regions are shown in blue and red, respectively, and the  \textcolor{black}{green} curve shows the marginal stability curve.}
\label{fig:translationgrowthrates_invertedproblem}
\end{figure}

In the translational mode ($m=1$), capillary stresses do not play a role, and the active and elastic growth rates are
\begin{align}
    \textcolor{black}{\Xi_{1}^{a}(\eta,\lambda)  = 3\eta \overline{A}_{1} }  , \qquad
   \textcolor{black}{ \Xi_{1}^{e}(\eta,\lambda)  = -\frac{6}{\eta}\overline{A}_{1}}.
\end{align}
These growth rates are plotted in Fig.~\ref{fig:translationgrowthrates_invertedproblem} vs $\eta$ and $\lambda$.\ Active stresses are destabilizing in extensile systems and stabilizing in contractile systems, whereas elasticity always resists translation. Both active and elastic effects are most pronounced for small drops and at low viscosity ratios. In extensile systems with finite elasticity, the stability of the system in the $(\eta,\lambda)$ parameter space is shown in Fig.~\ref{fig:translationgrowthrates_invertedproblem}(c), \textcolor{black}{where the destabilizing effect of active stresses manifests itself above a critical radius ratio whose value depends on $\xi$. {Contractile systems in the presence of elasticity are always stable to translation. } }

For deformation modes $(m\geq2)$, the capillary, active and elastic growth rates are expressed as
\begin{align}
    \Xi_{m}^{c}(\eta,\lambda) & = m(m^2-1) \left[\eta^{m-3}\overline{A}_{m}  +\eta^{m-1} \overline{C}_{m}\right], \label{eq:sigmac2} \\[3pt]
    \Xi_{m}^{a}(\eta,\lambda) & = \textcolor{black}{m\eta^{m-2}\left[2(m-1)  \overline{A}_{m}+ (2m+1)\eta^2\overline{C}_{m}\right]} ,\\[3pt]
    \Xi_{m}^{e}(\eta,\lambda) & = \textcolor{black}{m\eta^{m-4}\left[(m^3-2m^2-1)\overline{A}_{m}+ (m^3-2m^2-4m-1)\eta^{2}\overline{C}_{m}\right]}.
\end{align}
Equation~(\ref{eq:sigmac2}) for the capillary growth rate is the same as (\ref{eq:sigmac1}) and was already analyzed in Fig.~\ref{fig:deformationgrowthrates}(a) and (d).  

\begin{figure}[t]
\centering
\includegraphics[width=0.75\textwidth]{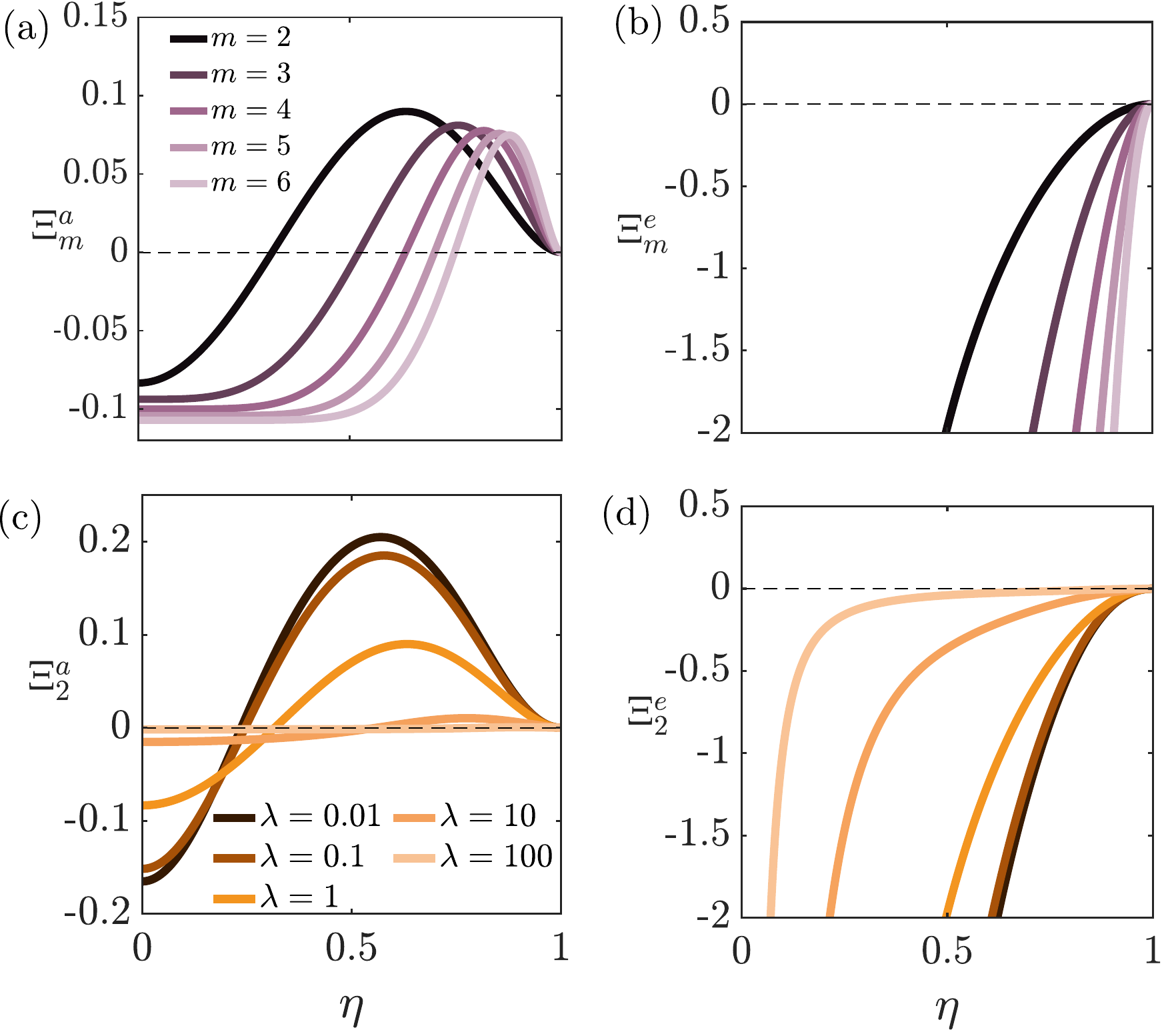}
\caption{(a) Active contribution $\Xi^a_m$, and (b) elastic contribution $\Xi^e_m$ to the growth rate $\Lambda_m$ for translational modes of increasing wavenumber $m$, plotted as functions of $\eta$ for $\lambda=1$. Panels (c) and (d) show the same functions for mode $m=2$ for different values of the viscosity ratio $\lambda$.}
\label{fig:deformationgrowthrates_invertedproblem}
\end{figure}

The variation of the active and elastic contributions to the growth rate $\Lambda_m$ for deformation modes of increasing wavenumber $m$ is shown in Fig.~\ref{fig:deformationgrowthrates_invertedproblem}(a,b). In extensile systems, activity is \textcolor{black}{stabilizing in small drops and destabilizing in larger drops, especially at low wavenumbers.} Elasticity, on the other hand, is \textcolor{black}{always} stabilizing and its effect is most pronounced in smaller drops at low wavenumbers. The effect of viscosity ratio is analyzed in Fig.~\ref{fig:deformationgrowthrates_invertedproblem}(c,d) for mode $m=2$, which shows that low viscosity drops are the most affected by both activity and elasticity.  


\section{Conclusions}\label{section5}

We have analyzed the linear stability of a passive viscous Newtonian drop immersed in an active nematic liquid crystal in two dimensions. In extensile systems, activity was found to stabilize the translational mode, but to destabilize deformation modes for small to intermediate drop sizes; these findings are reversed in contractile systems. Elasticity has a predominantly stabilizing effect, \textcolor{black}{except in the translational mode where it destabilizes drops of small to intermediate sizes.} We also considered the inverted problem of an active nematic drop surrounded by a viscous layer, where extensile stresses were found to be always destabilizing. When active, elastic, and capillary stresses act in concert, complex trends are predicted, with a non-trivial and non-intuitive dependence of the stability results on droplet size and viscosity ratio. \textcolor{black}{Note that we have analyzed the translation and deformation modes separately: in an experimental system, multiple modes will typically be excited simultaneously, and the mode subject to the largest unstable growth rate is expected to dominate the dynamics. }

Our model relied on several strong assumptions that may need to be relaxed or improved upon for comparison with specific experimental systems. Specifically, we considered a sharply aligned liquid crystal in the limit of strong nematic relaxation ($\mathrm{Pe}\rightarrow 0$) and with strong anchoring boundary conditions. A consequence of these choices is that the configuration of the nematic is fully determined by the instantaneous geometry of the system and is not directly affected by the viscous flow driven by the various stresses in the bulk of the nematic. If the assumption of strong elasticity is relaxed, additional instabilities can arise that involve spontaneous swirling flows, as has been observed in experiments \cite{woodhouse2012spontaneous,ray2023rectified} and other models \cite{furthauer2012taylor,theillard2017geometric}; the impact of such flows on the stability problem analyzed here in unclear. The potential emergence and role of topological defects is another question of interest: no defects were present in the model analyzed here (except for the $+1$ defect at the origin in the inverted problem of Sec.~\ref{section4}), but defects may arise in other geometries \cite{chandler2024active,neville2024controlling} and are likely to have an impact on the stability and dynamics. The interaction of multiple droplets suspended in an active nematic is also an open topic of interest. 

\textcolor{black}{Our theoretical model may be relevant to a wide range of biological systems. As an example, we highlight the dynamics and self-organization inside the cell nucleus, a highly heterogeneous confined environment involving the coexistence of transcriptionally active euchromatin, thought to behave as a nematic active fluid \cite{bruinsma2014,saintillan2018extensile,eshghi2021,mahajan2022euchromatin}, with various passive compartments such as transcriptionally silent heterochromatin domains, which are passive and viscoelastic \cite{eshghi2021}, as well as nucleoli, which behave as viscous liquid droplets \cite{caragine2018surface}. Experimental observations suggest the preferential placement of heterochromatin domains near the nuclear envelope, whereas nucleoli tend to be located in the bulk. While our understanding of the rheological behavior of these various phases and of their physical interactions is in its infancy, models such as ours may provide an useful tool for analyzing and explaining their spatiotemporal organization.}

\begin{acknowledgments}

D.S. acknowledges financial support from National Science Foundation Grant No.~DMS-2153520. 

\end{acknowledgments}

\renewcommand{\theequation}{A-\arabic{equation}}    
\setcounter{equation}{0}  
\section*{Appendix: Auxiliary Newtonian problem}

The Stokes equations (\ref{eq:recipflow})--(\ref{eq:recipflow2}) lead to the biharmonic equation $\nabla^4\psi = \nabla^4\overline{\psi}=0$ for the streamfunctions, which translate to an equidimensional differential equation for $G_{m}(r)$,
\begin{equation}
\frac{\text{d}^{4} G_{m}}{\text{d} r^{4}} + \frac{2}{r}\frac{\text{d}^{3} G_{m}}{\text{d} r^{3}} - \frac{2m^{2} +1}{r^{2}}\frac{\text{d}^{2} G_{m}}{\text{d} r^{2}} +  \frac{2m^{2} +1}{r^{3}}\frac{\text{d} G_{m}}{\text{d} r} + \frac{m^{4} -4m^{2}}{r^{4}}G_m = 0,   \label{eq:odeG}
\end{equation}
and an identical equation for $\overline{G}_{m}(r)$. The boundary conditions on $G_{m}$ and $\overline{G}_{m}$ are
\begin{itemize}
\item the no-slip condition at $r=1$:
\begin{equation}
G_{m}(1) = G_{m}'(1) = 0,   \label{eq:noslipG}
\end{equation}
\item the continuity of velocities at $r=\eta$:
\begin{equation}
G_{m}(\eta) = \overline{G}_{m}(\eta), \qquad G_{m}'(\eta)=\overline{G}_{m}'(\eta),
\end{equation}
\item the stress balance of Eq.~(\ref{eq:stressjump}) in the radial and azimuthal directions:
\begin{align}
\eta^3 G_{m}''' +\eta^2 G_{m}'' -(1+3m^2)\eta G_{m}' + 4m^2 G_{m} &= \lambda \left[\eta^3 \overline{G}_{m}''' +\eta^2 \overline{G}_{m}'' -(1+3m^2)\eta \overline{G}_{m}' + 4m^2 \overline{G}_{m}\right]- m \eta^2 ,  \\
\eta^2 G_{m}''-\eta G_{m}' +m^2 G_{m} &= \lambda \left[\eta^2 \overline{G}_{m}''-\eta \overline{G}_{m}' +m^2 \overline{G}_{m}\right] \qquad \mathrm{at}\,\,\, r=\eta,   \label{eq:stressjumpG}
\end{align}
\item boundedness of the velocity at the origin $r=0$.
\end{itemize}
These boundary conditions can be used to solve for the eight integration constants obtained when integrating (\ref{eq:odeG}) in both fluid domains. Solution details differ for $m=1$ (translation mode) and $m>1$ (deformation modes).  

\subsection*{Translation mode $(m=1)$}

When $m=1$, Eq.~(\ref{eq:odeG}) for $G_{m}(r)$ admits the solution
\begin{equation}
G_{1}(r) = A_{1}r^{3} +  B_{1}r\ln r + C_{1}r + \frac{D_{1}}{r},  \label{eq:Gtrans}
\end{equation}
where $A_1$, $B_1$, $C_1$ and $D_1$ are dimensionless functions of the viscosity ratio $\lambda$ and radius ratio $\eta$. A similar expression holds for $\overline{G}_{m}(r)$, where $\overline{B}_1=\overline{D}_1=0$ to ensure boundedness of the velocity at the origin. Applying boundary conditions (\ref{eq:noslipG})--(\ref{eq:stressjumpG}) provides six linear equations for the six unknown coefficients, with the solutions
\begin{eqnarray}
&\displaystyle A_{1}= -\frac{\eta}{8}\frac{(1-\eta^2)\lambda+1}{(1-\eta^4)\lambda +1+\eta^4}\,,   \qquad &B_1 = \frac{\eta}{4}\,,  \label{eq:A1} \\ 
& \displaystyle C_1 = \frac{\eta}{8}\frac{(1-\eta^2)^2 \lambda +1-\eta^4}{(1-\eta^4)\lambda +1+\eta^4}\,, \qquad &D_1 =\frac{\eta^3}{8}\frac{(1-\eta^2)\lambda +\eta^2}{(1-\eta^4)\lambda +1+\eta^4}\,,   \\
&\displaystyle \overline{A}_1 = \frac{1}{8\eta}\frac{(1-\eta^2)^2}{(1-\eta^4)\lambda +1+\eta^4}\,,\qquad &\overline{C}_1 = \frac{\eta}{4}\left[\frac{(1-\eta^2)^2 \lambda +\eta^2(1-\eta^2)}{(1-\eta^4)\lambda +1+\eta^4}+\ln\eta  \right].\label{eq:A1bar}
\end{eqnarray}

\subsection*{Deformation modes $(m>1)$}

When $m>1$, the expression for $G_{m}(r)$ reads
\begin{equation}
G_{m}(r)=A_{m}r^{m} + \frac{B_{m}}{r^{m}} +  C_{m}r^{m+2} + \frac{D_{m}}{r^{m-2}},  \label{eq:Gdef}
\end{equation}
with a similar expression for $\overline{G}_{m}(r)$ in which $\overline{B}_m=\overline{D}_m=0$ to ensure boundedness at $r=0$. As for $m=1$, applying the boundary conditions yields six equations for the coefficients, which are obtained as
\begin{align}
A_m&=\frac{\mathcal{K}_m}{m-1}\left\{ [m(\eta^2-1)-\eta^2](\lambda+1)+\eta^{2m+2}(\lambda-1) \right\},  \label{eq:Am}\\
B_m&=-\frac{\mathcal{K}_m}{m+1}\left\{ \eta^2 (\lambda+1)+[-(m+1)\eta^{2m+2}+m\eta^{2m}](\lambda-1)  \right\} , \\
C_m&=-\frac{\mathcal{K}_m}{m+1}\left\{ [m(\eta^2-1)-1 ](\lambda+1) + \eta^{2m}(\lambda-1)   \right\},  \\
D_m&=\frac{\mathcal{K}_m}{m-1}\left\{ (\lambda+1)-[m\eta^{2m+2}-(m-1)\eta^{2m}](\lambda-1)  \right\},
\end{align}
and
\begin{align}
\overline{A}_m &=\frac{\mathcal{K}_m}{m-1} \left\{ \eta^{2-2m}(\lambda+1)+[\eta^{2m+2}-m^2\eta^4  ](\lambda-1) +2(m-1)\eta^2 [m(\lambda-1)+\lambda] -m [m(\lambda-1)+2] \right\},   \\
\overline{C}_m &=- \frac{\mathcal{K}_m}{m+1}  \left\{ \eta^{-2m} (\lambda+1)+[\eta^{2m}-m^2\eta^{-2}](\lambda-1) + 2(m+1)[m(\lambda-1)-\lambda]  -m \eta^2[m(\lambda-1)-2] \right\},  \label{eq:Cmbar}
\end{align}
where the prefactor $\mathcal{K}_m$ is given by
\begin{align}
\mathcal{K}_m=\frac{1}{4}\eta^{m+2}\left\{[m^2 \eta^{2m+4}-2(m^2-1)\eta^{2m+2}+m^2\eta^{2m}](\lambda^2-1)-\eta^{4m+2}(\lambda-1)^2 -\eta^2(\lambda+1)^2\right\}^{-1}. 
\end{align}

\bibliography{references}

\end{document}